Quantum Analogical Modeling:
A General Quantum Computing Algorithm for Predicting Language Behavior

Royal Skousen <royal_skousen@byu.edu>
Department of Linguistics, Brigham Young University, Provo, Utah 84602 USA

18 October 2005

*Abstract*

This paper proposes a general quantum algorithm that can be applied to any classical computer program. Each computational step is written using reversible operators, but the operators remain classical in that the qubits take on values of only zero and one. This classical restriction on the quantum states allows the copying of qubits, a necessary requirement for doing general classical computation. Parallel processing of the quantum algorithm proceeds because of the superpositioning of qubits, the only aspect of the algorithm that is strictly quantum mechanical. Measurement of the system collapses the superposition, leaving only one state that can be observed. In most instances, the loss of information as a result of measurement would be unacceptable. But the linguistically motivated theory of Analogical Modeling (AM) proposes that the probabilistic nature of language behavior can be accurately modeled in terms of the simultaneous analysis of all possible contexts (referred to as supracontexts) providing one selects a single supracontext from those supracontexts that are homogeneous in behavior (namely, supracontexts that allow no increase in uncertainty). The amplitude for each homogeneous supracontext is proportional to its frequency of occurrence, with the result that the probability of selecting one particular supracontext to predict the behavior of the system is proportional to the square of its frequency.

Part 1: The Quantum Mechanical Properties of Analogical Modeling

1.1 *Introduction*

Analogical Modeling (AM) is a general theory for predicting behavior. It can also be considered a system of classification or categorization according to a particular set of outcomes. Predictions are directly based on a data set of exemplars. These exemplars give the outcome for various configurations of variables, which may be structured in different ways (such as strings or trees). A common method in AM is to define the variables so that there is no inherent structure or relationships between the variables (that is, each variable is defined independently of all the other variables). In this case, the variables can be considered a vector of features. In the data set, each feature vector is assigned an outcome vector. The data set is used to predict the outcome vector for a test set of various feature vectors for which no outcome vector has been assigned (or if one has been assigned, it is ignored). In AM the resulting predictions are not based on any learning stage for which the data set has been analyzed in advance in order to discover various kinds of potential relationships between the feature vectors and their associated outcome vectors. Neural nets, decision trees, and statistical analyses that determine the significance of the features in predicting the outcomes all rely on first learning something about the data set and then using that information to make predictions. AM, on the other hand, directly uses the data set to make a prediction for each specific feature vector in the test set.

   The basic theory of AM was developed from 1979-1987 and works from the hypothesis that in trying to predict the outcome (or behavior) for a vector of features, we consider all possible combinations of those features. Using a simple quadratic measure of uncertainty (not the traditional logarithmic one of information theory), we select those combinations of features that never permit any increase in the uncertainty. Those combinations that increase the uncertainty are referred to as heterogeneous and are eliminated from the analysis.

Another way to look at AM is to view each combination of features and its predicted outcome as a rule that maps from the feature combination to the outcome. The homogeneous combinations can be considered "true rules", the heterogeneous ones as "false rules". In other words, AM uses only the true rules; the false rules are ignored. Given that we have determined the true rules, the question then becomes: What are the chances of using a particular true rule to predict the outcome? The false rules, of course, are all assigned a probability of zero. Among the conceptual possibilities for assigning a probability to a true rule are the following: (1) each rule is equally probable; (2) the probability is proportional to the frequency of the rule in the data; (3) the probability is proportional to the square of the frequency of the rule (Skousen 1992:8-9). Over time, it has become clear that the third choice is the simplest and most natural since it directly uses the same quadratic measure of uncertainty that is already needed to determine which rules are true (that is, which feature combinations are homogeneous in behavior). Moreover, the third choice has provided the most accurate results in predicting language behavior, including the appropriate degree of fuzziness that occurs at the boundaries of language behavior.

AM has shown considerable success in explaining actual language behavior, beginning with Royal Skousen's description of the indefinite article in English (Skousen 1989:54-59 as well as Skousen 2003) and the past tense in Finnish (Skousen 1989:101-136), followed by Bruce Derwing and Royal Skousen's work on the past tense in English (Derwing & Skousen 1994), David Eddington on various problems in Spanish morphology (such as stress assignment in Eddington 2000a, diminutive formation in Eddington 2002a, gender assignment in Eddington 2002b, and nominalization in Eddington 2004:83-98), and Harald Baayen and his colleagues in the Netherlands on various aspects of Dutch morphology (see, for instance, Krott, Schreuder & Baayen 2001 and Ernestus & Baayen 2003). Steve Chandler has provided a thorough comparison of AM with connectionist models of language as well as with a number of competing instance-based models. Chandler has shown how AM, a single-route approach to language description (that is, AM has a single conceptual mechanism), can readily handle various experimental results that were earlier claimed to be possible only in dual-route approaches to language; moreover, Chandler has found evidence from various psycholinguistic results that only AM seems capable of explaining (Chandler 2002). Similarly, Eddington 2000b has provided evidence that single-route exemplar-based approaches can accurately model speakers' abilities to predict both regular and irregular forms in language. For additional information about AM, see the Analogical Modeling website at <http://humanities.byu.edu/am/>.

One interesting aspect of the theory of AM is that it has never undergone any fundamental revisions from its final conceptualization in 1987. The basic idea has remained constant: (1) all true rules (that is, homogeneous combinations of features, structured or unstructured) are possible, and (2) the probability of using a particular true rule to predict the outcome is proportional to the square of its frequency.

Early on it was realized that AM involved an exponential explosion. For every feature variable that was added to an analysis, there was basically a doubling in both the hardware requirements (memory) as well as the running time for a standard sequentially programmed computer to determine which feature combinations are homogeneous. In the simplest case, when there are $n$ features all independent of each other, the total number of possible combinations that must be considered is $2^n$. At a workshop held at Corsendonk, Belgium, late in 1997, the urgency of solving this exponential explosion (from at least a theoretical standpoint, if not a pragmatic one) became very clear (Round Table on Algorithms for Memory-Based Language Processing, Corsendonk, Turnhout, Belgium, 12-13 December 1997, organized by Walter Daelemans and Steven Gillis).

A little more than two years later it was discovered that the problem of exponentiality could be theoretically solved if AM were treated as a quantum mechanical problem for which the following would hold:

> (1) all possible rules exist in a superposition;
>
> (2) the system evolves so that
>> (a) the amplitude of every heterogeneous rule equals zero, while
>>
>> (b) the amplitude of each homogeneous rule equals its relative frequency of occurrence;
>
> (3) measurement or observation reduces the superposition to a single rule, with the probability of it being selected equal to its amplitude squared (that is, equal to its relative frequency squared).

In 2000, I was able to demonstrate in general, but not in detail, how classical reversible programming combined with the superpositioning of quantum mechanics could be used to solve this outstanding problem in AM exponentiality (posted as a preprint on <http://www.arXiv.org> on 28 August 2000; published in Skousen, Lonsdale & Parkinson 2002).

The purpose of this paper is to provide the specifics of how quantum computing would do Analogical Modeling. This paper is divided into two main parts. In the first, I outline how AM works. In particular, I define a specific sample data set that is used throughout the paper, especially in part 2 where I specifically show how the quantum computing version of AM solves this simple example. I will refer to the quantum mechanical reinterpretation of AM as QAM (namely, Quantum Analogical Modeling). In discussing QAM, the following properties will be noted:

(1) There is only one essential quantum concept in QAM: namely, the *superposition* of all possible rules; technically, the contextual specification for each rule is referred to as a supracontext (that is, we have a superposition of all the supracontexts).

(2) The actual operators used to control the evolution of the superposition are *classical reversible operators*.

(3) The operators must be written so they are applicable to every supracontext in the superposition; there is no conditional application, only *universal application.*

(4) Since the operators are classical reversible operators, the *only possible states* are 0 and 1. We avoid the full range of possibilities defined by generalized qubits; we use qubits in QAM, but we make sure that these qubits effectively take only the states 0 and 1.

(5) The result of using only the states 0 and 1 means that *qubits can be copied* and that fanout is therefore possible. Copying qubits allows us to replicate any possible classical reversible operator; on the other hand, copying is not possible for generalized qubits.

(6) Heterogeneity is a specific property of supracontexts that can be identified and derived by classical reversible operators; in QAM the superposition evolves so that we are able to identify the heterogeneous supracontexts and *set their amplitude to zero.*

(7) For each homogeneous supracontext, the final amplitude is *proportional to the frequency of occurrence* for that supracontext.

(8) From a rule perspective, quantum mechanical measurement or observation *collapses the superposition* of supracontexts (all possible rules) to a single supracontext (only one of the rules), which is then used to predict the outcome.

(9) From the exemplar perspective, measurement in QAM occurs by *randomly selecting a pointer* to any one of the data items in any occurring homogeneous supracontext; empty supracontexts have no data items and therefore no pointers; the pointers between the data items in heterogeneous supracontexts are effectively eliminated (that is, rendered unobservable).

(10) This one-stage system of measurement in exemplar-based QAM *avoids the traditional two-stage measurement system* in quantum mechanics, yet both give exactly the same result: (a) the amplitude is equal to the relative frequency of occurrence for every homogeneous supracontext, but equal to zero for every heterogeneous supracontext; (b) the amplitude squared gives the probability of using a particular supracontext to predict the outcome.

(11) The *directional pointers* connecting each pair of data items are used to determine the uncertainty and consequently the heterogeneity of any given supracontext in the superposition. These same pointers are used to observe the system; since the number of pointers is equal to the square of the number of data items, observation automatically leads to the squaring of the amplitude.

(12) QAM is *a general quantum computational algorithm* that categorizes and predicts behavior; it is based on the notion that supracontexts can be classified according to a specific definition of heterogeneity. Classical reversible programming can be used to identify other properties for superpositions of supracontexts, but whether other properties will serve useful roles in predicting behavior (or anything else) is an open issue.

Three kinds of quantum algorithms have been recognized as computational improvements over classical algorithms: (1) those based on Fourier transforms, such as Peter Shor's algorithm for determining the prime factors of an integer, (2) search algorithms dealing with finding a specific item in an unordered list, with the first example discovered by Lov Grover, and (3) quantum simulation of the behavior of *n* qubits for a superposition of $2^n$ cases. This third possibility was originally discussed by Richard Feynman (1982, reprinted in Hey 1999:133-153). In discussing this third possibility, Nielsen and Chuang (2000:39) warn that in measuring the superposition of $2^n$ cases, we can observe the information contained in only one of those cases:

> A significant caveat is that even though a quantum computer can simulate many quantum systems far more efficiently than a classical computer, this does not mean that the fast simulation will allow the desired information about the quantum system to be obtained. When measured, a *kn* qubit simulation will collapse into a definite state, giving only *kn* bits of information; the $c^n$ bits of 'hidden information' in the wavefunction is not entirely accessible.

Analogical Modeling, it turns out, proposes such a collapse into a single definite state, yet that loss of information actually predicts how language behaves – and at the appropriate level of uncertainty.

1.2 *A Basic Example of Analogical Modeling*

Throughout this paper, I will be using a simple example to show how AM (and QAM) works. This example has a data set defined by three independent categorical variables. These variables, referred to as features, will be used to predict whether a speaker uses a form *x* or *y*. (This example is based on a term of address in Arabic; for discussion, see Skousen 1989:97-100.) The first feature variable records the age of the speaker, the second whether the speaker is male or female, and the third feature variable notes the relationship between the speaker and the person being addressed:

|  |  |  |
|---|---|---|
| (1) age of speaker | c | **c**hild |
|  | g | **g**rown-up (but not old) |
|  | o | **o**ld |
| (2) gender of speaker | m | **m**ale |
|  | f | **f**emale |
| (3) relationship with addressee | n | **n**uclear family |
|  | r | **r**elative |
|  | a | **a**cquaintance |
|  | s | **s**tranger |

The outcome to be predicted is a term of address that means 'my brother'. There are two possibilities:

|  |  |  |
|---|---|---|
| outcome | x | ya'a**xi** |
|  | y | yaxuu**ya** |

The three feature variables and the one outcome variable are listed here only to show how AM works. We ignore here the possibility that the first and third of these feature variables could be recast as scalars rather than distinct nonscalar categories.

Our simple data set will contain six exemplars:

| variable 1 | variable 2 | variable 3 | outcome |
|---|---|---|---|
| *speaker's age* | *speaker's gender* | *addressee's relationship to speaker* | *term of address* |
| **o**ld | **m**ale | **s**tranger | yaxuu**ya** |
| **g**rown-up | **f**emale | **a**cquaintance | ya'a**xi** |
| **c**hild | **m**ale | **s**tranger | ya'a**xi** |
| **c**hild | **m**ale | **a**cquaintance | ya'a**xi** |
| **o**ld | **m**ale | **n**uclear family | ya'a**xi** |
| **g**rown-up | **f**emale | **r**elative | ya'a**xi** |

Our task will be to predict the behavior of various items in the test set, in particular the following test item:

**o**ld             **m**ale             **a**cquaintance             ya'a**x**i *or* yaxuu**y**a ?

Using one-letter symbols (marked above in bold) for the feature variables and the outcome variable, we have the following data set and test set:

*data set*

| | | | |
|---|---|---|---|
| o | m | s | y |
| g | f | a | x |
| c | m | s | x |
| c | m | a | x |
| o | m | n | x |
| g | f | r | x |

*test set*

| | | |
|---|---|---|
| o | m | a | x *or* y ? |

## 1.3 *An Exemplar-Based Determination of Heterogeneity*

A specific feature vector in the test set is typically referred to as a given context. In general, our task is to predict the probability that the outcome for a given context will be *x* or *y*. In this particular example, the test set has only one feature vector, *oma* (that is, an old male speaking with a male acquaintance). We want to predict whether such a speaker will use the *x* form or the *y* form (that is, either *ya'axi* or *yaxuuya* for 'my brother'). Going over the available data (the data set), we find the following instances for all possible combinations of the feature variables *o, m,* and *a*. In the following list, the short dash (-) is used as a wild card symbol; there is no restriction on the feature value for any variable represented as such:

| | | | | | | | | |
|---|---|---|---|---|---|---|---|---|
| o | m | a | no instances | — | | | | |
| o | m | - | two instances | o | m | s | → | y |
| | | | | o | m | n | → | x |
| o | - | a | no instances | — | | | | |
| - | m | a | one instance | c | m | a | → | x |
| o | - | - | two instances | o | m | s | → | y |
| | | | | o | m | n | → | x |
| - | m | - | four instances | o | m | s | → | y |
| | | | | c | m | s | → | x |
| | | | | c | m | a | → | x |
| | | | | o | m | n | → | x |

|  |  |  | two instances | g | f | a | → | x |
|---|---|---|---|---|---|---|---|---|
| - | - | a |  | c | m | a | → | x |
|  |  |  |  |  |  |  |  |  |
| - | - | - | all instances | o | m | s | → | y |
|  |  |  |  | g | f | a | → | x |
|  |  |  |  | c | m | s | → | x |
|  |  |  |  | c | m | a | → | x |
|  |  |  |  | o | m | n | → | x |
|  |  |  |  | g | f | r | → | x |

We refer to each of these possible combinations as a supracontext for the given context *oma*. If we have $n$ independent feature variables in a given context, there will be $2^n$ supracontexts for that given context. One supracontext, *oma* (the one that includes all $n$ feature variables), will be identical to the given context; there will also be one completely general supracontext, ---, that ignores all the feature variables and therefore includes all the items in the data set. We can think of each supracontext as a distinct set of criteria for predicting the behavior (that is, the appropriate outcome) for the given context. The use of the short dash (-) for a particular feature variable in a supracontext means that that feature variable is being ignored for that supracontext. Thus --*a* means that the first and second feature variables are being ignored; we will look for exemplars for which the third feature variable is *a* (the male being addressed is an acquaintance).

One important aspect of Analogical Modeling is that the predictions are not done in terms of probabilities but in terms of specific instances or exemplars. We do NOT calculate a probability from the instances and use that probability to predict an outcome. Instead, we randomly select one of the instances and predict behavior according to the outcome assigned to that instance.

1.4  *Measuring Uncertainty*

The key to Analogical Modeling is that we accept only those supracontexts for predicting behavior that never allow the uncertainty to increase. In AM, uncertainty is defined in a very simple manner.

The normal approach for measuring the uncertainty of rule systems is Shannon's "information", the logarithmic measure more commonly known as the entropy *H*. Although this measure can be given a natural interpretation (as the average number of yes-no questions needed to determine the outcome of a supracontextual specification), it has some disadvantages: (1) entropy is based on the notion that one gets an unlimited number of chances to discover the correct outcome; (2) the entropy for continuous probabilistic distributions is infinite; even the entropy density is infinite for continuous distributions, and an unmotivated definition for entropy density must be devised, one that sometimes gives negative measures of entropy density (see the discussion in Skousen 1992:89-91).

A more plausible and simpler method for measuring uncertainty is a quadratic one, the disagreement *Q*. This measure has a natural interpretation: it represents the probability that two randomly chosen instances of a supracontext disagree in outcome. The disagreement is based on the plausible restriction that one gets a single chance to guess the correct outcome rather than an unlimited number of guesses. Moreover, the disagreement density exists (and is positive finite) for virtually all continuous probabilistic distributions. In fact, the disagreement density can be used to measure the uncertainty of continuous distributions for which the variance (the traditional measure of dispersion) is undefined (see Skousen 1992:73-84).

The key to measuring $Q$ is the notion of disagreement. Suppose we have a particular supracontext with $n$ data items, and for each outcome $\omega_i$ we have $n_i$ data items with that outcome. For each data item we compare its outcome with the outcome of every data item, including itself. If a given pair of outcomes are identical, we have an agreement; if they differ, we have a disagreement. We get the following for the number of agreements and disagreements:

| | |
|---|---|
| number of agreements for outcome $\omega_i$ | $n_i^2$ |
| number of disagreements for outcome $\omega_i$ | $n_i(n-n_i)$ |
| | |
| total number of agreements | $\Sigma\, n_i^2$ |
| total number of disagreements | $\Sigma\, n_i(n-n_i) \;=\; n^2 - \Sigma\, n_i^2$ |
| total number of pairings | $n^2$ |

The uncertainty $Q$ is the fraction of disagreement:

$$Q \;=\; \Sigma\, n_i(n-n_i)/n^2 \;=\; (n^2 - \Sigma\, n_i^2)/n^2 \;=\; 1 - \Sigma\,(n_i/n)^2$$

The task of Analogical Modeling is to find those supracontexts for which the disagreement is minimized. If the subcontextual partition for the supracontext has fewer disagreements, then the supracontext is heterogeneous; but if the subcontextual partition and its supracontext have the same number of disagreements, then the supracontext is homogeneous.

1.5 *Determining the Homogeneity of Supracontexts*

In this section, we consider the specifics of how to determine the homogeneity for each supracontext in terms of how the individual data occurrences behave. We show that homogeneity can be determined by observing the behavior of the data items found in each supracontext alone, without comparing the behavior of one supracontext to any other supracontext.

As before, we have the given context *oma*. For each data item in the data set, we determine its subcontext and its outcome:

*given context*
o   m   a

| data item | subcontext | outcome |
|---|---|---|
| o   m   s   y | omā | y |
| g   f   a   x | ōma | x |
| c   m   s   x | ōmā | x |
| c   m   a   x | ōma | x |
| o   m   n   x | omā | x |
| g   f   r   x | ōmā | x |

The subcontexts are defined with respect to the given context. For each data item, we compare the variables of that data item with the given context. For instance, when compared with the given context *oma*, the data item *cms* differs in the first and third variables but is identical in the second variable. We represent this comparison as *ōmā*, where the macron over a variable means 'not'.

The given context also defines all the possible supracontexts. For each supracontext, we list the specific data items contained within that supracontext; and for each of those specific data items, we identify its subcontext and its outcome:

| supracontext | specific data items | subcontext | outcome |
|---|---|---|---|
| o  m  a | — | | |
| o  m  - | o  m  s  y | omā | y |
|         | o  m  n  x | omā | x |
| o  -  a | — | | |
| -  m  a | c  m  a  x | ōma | x |
| o  -  - | o  m  s  y | omā | y |
|         | o  m  n  x | omā | x |
| -  m  - | o  m  s  y | omā | y |
|         | c  m  s  x | ōmā | x |
|         | c  m  a  x | ōma | x |
|         | o  m  n  x | omā | x |
| -  -  a | g  f  a  x | ōm̄a | x |
|         | c  m  a  x | ōma | x |
| -  -  - | o  m  s  y | omā | y |
|         | g  f  a  x | ōm̄a | x |
|         | c  m  s  x | ōmā | x |
|         | c  m  a  x | ōma | x |
|         | o  m  n  x | omā | x |
|         | g  f  r  x | ōm̄ā | x |

Now for each supracontext, we determine whether the list of contained data items manifests a single subcontext or a single outcome (the *or* here is inclusive). If so, the supracontext is homogeneous. But if there is more than one subcontext listed and more than one outcome listed, the supracontext is heterogeneous:

| supracontext | distinguishable subcontexts | | distinguishable outcomes | |
|---|---|---|---|---|
| o  m  a | – | 0 | – | 0 |
| o  m  - | omā | 1 | x, y | 2 |
| o  -  a | – | 0 | – | 0 |
| -  m  a | ōma | 1 | x | 1 |
| o  -  - | omā | 1 | x, y | 2 |
| -  m  - | omā, ōmā, ōma | 3 | x, y | 2 |
| -  -  a | ōm̄a, ōma | 2 | x | 1 |
| -  -  - | omā, ōm̄a, ōmā, ōma, ōm̄ā | 5 | x, y | 2 |

We don't actually need to list the distinguishable subcontexts or the distinguishable outcomes – or to even count them. Rather, we simply look for plurality of subcontexts and plurality of outcomes. If we get plurality for both subcontexts and outcomes, then the supracontext is heterogeneous; otherwise, it is homogeneous:

| supracontext | | | plurality of subcontexts | plurality of outcomes | heterogeneity |
|---|---|---|---|---|---|
| o | m | a | no | no | no |
| o | m | - | no | yes | no |
| o | - | a | no | no | no |
| - | m | a | no | no | no |
| o | - | - | no | yes | no |
| - | m | - | yes | yes | yes |
| - | - | a | yes | no | no |
| - | - | - | yes | yes | yes |

Each data item within a supracontext is connected by means of a directional pointer to every other data item (including itself) within the supracontext, as described in section 1.4. Pointers are accessible only within homogeneous supracontexts. We randomly select one of the accessible pointers, look at what data item it is pointing to, identify the outcome for that data item, and use that outcome to predict the behavior for the given context. Thus we end up with the following results:

| supracontext | | | homogeneity | accessible occurrences | | pointers to | |
|---|---|---|---|---|---|---|---|
| | | | | outcome x | outcome y | outcome x | outcome y |
| o | m | a | yes | 0 | 0 | 0 | 0 |
| o | m | - | yes | 1 | 1 | 2 | 2 |
| o | - | a | yes | 0 | 0 | 0 | 0 |
| - | m | a | yes | 1 | 0 | 1 | 0 |
| o | - | - | yes | 1 | 1 | 2 | 2 |
| - | m | - | no | 0 | 0 | 0 | 0 |
| - | - | a | yes | 2 | 0 | 4 | 0 |
| - | - | - | no | 0 | 0 | 0 | 0 |
| | | | | | *pointer totals* | 9 | 4 |

It is important to note that in AM we do not have to collapse to one supracontext and then randomly select the outcome for that supracontext. From the entire superposition of homogeneous supracontexts, we can randomly select one of the 13 ( = 9 + 4 ) available pointers; the outcome that it points to is the predicted outcome for that measurement of the system. One can, of course, make predictions using the traditional two-step method of quantum mechanical observation, but it is important to note that it is not necessary in AM.

## 1.6 Predictions for an Expanded Test Set

In the previous sections, we have used AM to predict the behavior for only one test item – namely, the the given context *oma*. In the following discussion, we consider the predicted behavior for the following additional test items: *oms, gms, ofs, ofn, gmr, cfs,* and *cfa*. Only the first one, *oms*, occurs in the data set as one of the six data items. In fact, this is the only data item that takes the outcome *y*; all the others take the outcome *x*. This basically means that the data set exemplifies regular/exceptional behavior: namely, every data item takes the outcome *x* (the regular behavior) except for *oms*, which takes the outcome *y* (the exceptional behavior). In Analogical Modeling, test items close to an exceptional data item have some chance of behaving like the exceptional item, but the further the test item is from the exceptional item, the more regular the behavior. Normally the regular behavior dominates unless the test item is the exceptional item itself. Even when the test item differs in only one feature variable from the exceptional item, the regular behavior will dominate (which is one reason why the outcome *x* is called regular).

These properties are exemplified in the AM predictions for the expanded test set as listed above. For each test item, we follow the AM procedure, treating each test item as a given context and determining its homogeneous supracontexts in accord with the data set. In the following chart, I categorize the test items according to their distance from the exceptional data item *oms* (that is, according to the number of differing feature variables). The only test item that actually occurs as a data item is the exceptionally behaving *oms*, which means that for it AM predicts the exceptional behavior; none of the other test items are found in the data set, yet AM readily makes an appropriate prediction according to the exceptional/regular behavior of the data set:

| *distance* | *given context* | *probability of prediction* | |
|---|---|---|---|
| | | *regular* x | *exceptional* y |
| 0 | oms | 0.000 (0/2) | 1.000 (2/2) |
| 1 | oma | 0.692 (9/13) | 0.308 (4/13) |
| | gms | 0.667 (8/12) | 0.333 (4/12) |
| | ofs | 0.800 (4/5) | 0.200 (1/5) |
| 2 | ofn | 1.000 (6/6) | 0.000 (0/6) |
| | gmr | 0.818 (18/22) | 0.182 (4/22) |
| | cfs | 1.000 (9/9) | 0.000 (0/9) |
| 3 | cfa | 1.000 (14/14) | 0.000 (0/14) |

## 1.7 Statistical Overview of Heterogeneity

There is a single statistical test for determining the heterogeneity of a supracontext: namely, we compare the number of disagreements within the supracontext with the sum of all the disagreements in the subcontexts for that supracontext (the formulas for determining the number of disagreements can be found in section 1.4). This is equivalent to comparing the uncertainty between the supracontext and the full subcontextual partition for that supracontext (as defined by the given context):

| supracontext | data items | disagreements | subcontextual partition | data items | disagreements |
|---|---|---|---|---|---|
| o  m  a | – | 0 | o  m  a | – | 0 |
| o  m  - | oms**y** omn**x** | 2 | o  m  a<br>o  m  ā | –<br>oms**y** omn**x** | 0<br>2 |
| o  -  a | – | 0 | o  m  a<br>o  m̄  a | –<br>– | 0<br>0 |
| -  m  a | cma**x** | 0 | o  m  a<br>ō  m  a | –<br>cma**x** | 0<br>0 |
| o  -  - | oms**y** omn**x** | 2 | o  m  a<br>o  m  ā<br>o  m̄  a<br>o  m̄  ā | –<br>oms**y** omn**x**<br>–<br>– | 0<br>2<br>0<br>0 |
| -  m  - | oms**y** cms**x** cma**x** omn**x** | 6 | o  m  a<br>o  m  ā<br>ō  m  a<br>ō  m  ā | –<br>oms**y** omn**x**<br>cma**x**<br>cms**x** | 0<br>2<br>0<br>0 |
| -  -  a | gfa**x** cma**x** | 0 | o  m  a<br>o  m̄  a<br>ō  m  a<br>ō  m̄  a | –<br>–<br>cma**x**<br>gfa**x** | 0<br>0<br>0<br>0 |
| -  -  - | oms**y** gfa**x** cms**x** cma**x** omn**x** gfr**x** | 10 | o  m  a<br>o  m  ā<br>o  m̄  a<br>o  m̄  ā<br>ō  m  a<br>ō  m  ā<br>ō  m̄  a<br>ō  m̄  ā | –<br>oms**y** omn**x**<br>–<br>–<br>cma**x**<br>cms**x**<br>gfa**x**<br>gfr**x** | 0<br>2<br>0<br>0<br>0<br>0<br>0<br>0 |

If the number of disagreements is the same for the supracontext and its subcontextual partition, then the supracontext is homogeneous; if the number of disagreements is less in the subcontextual partition, then the supracontext is heterogeneous:

| supracontext | disagreements | sum of subcontextual disagreements | difference | homogeneity |
|---|---|---|---|---|
| o m a | 0 | 0 | 0 | yes |
| o m - | 2 | 2 | 0 | yes |
| o - a | 0 | 0 | 0 | yes |
| - m a | 0 | 0 | 0 | yes |
| o - - | 2 | 2 | 0 | yes |
| - m - | 6 | 2 | 4 | no |
| - - a | 0 | 0 | 0 | yes |
| - - - | 10 | 2 | 8 | no |

From this general definition of homogeneity, we derive the following two properties:

(1) *If a supracontext is deterministic in its outcomes, then the supracontext is homogeneous.*

A deterministic supracontext has no disagreements; no matter what the subcontexts are, they too will be deterministic (or empty) and will therefore show no disagreements; thus every deterministic supracontext is homogeneous.

(2) *If a supracontext is deterministic in its subcontexts, then the supracontext is homogeneous.*

If a supracontext is nondeterministic in outcome, then that supracontext will be homogeneous only if all the data items can be found in a single subcontext. Suppose the subcontextual partition separates off a data item with outcome *x* from a data item with outcome *y*. Then the two disagreements between those two data items will be eliminated in the subcontextual partition and the number of disagreements will therefore decrease:

| supracontext | disagreements | subcontextual partition | sum of disagreements |
|---|---|---|---|
| x y | 2 | x \| y | 0 |

If the subcontextual partition keeps the *x* and *y* outcomes together, then there is no decrease:

| supracontext | disagreements | subcontextual partition | sum of disagreements |
|---|---|---|---|
| x y | 2 | x y \| | 2 |

But the addition of any third data item in a different subcontext will lead to a decrease (no matter what its outcome, either *x* or *y* or some other outcome *z*):

| supracontext | disagreements | subcontextual partition | sum of disagreements |
|---|---|---|---|
| x y x | 4 | x y \| x | 2 |
| x y y | 4 | x y \| y | 2 |
| x y z | 6 | x y \| z | 2 |

The only time the number of disagreements remains unchanged is when all these data items (with differing outcomes) occur in the same subcontext:

| supracontext | disagreements | subcontextual partition | sum of disagreements |
|---|---|---|---|
| x y x | 4 | x y x \| | 4 |
| x y y | 4 | x y y \| | 4 |
| x y z | 6 | x y z \| | 6 |

If a supracontext is nondeterministic in both outcomes and subcontexts, then it will be heterogeneous.

One striking property of Analogical Modeling is that a more general nondeterministic supracontext will be homogeneous only if it has one occurring subcontext – and that subcontext would therefore be a redundant version of the supracontext. To consider this property, suppose we have a simple data set with the following eight data items: *oqx, oqy, orx, ory, pqx, pqy, prx,* and *pry* – that is, the first feature variable can be either *o* or *p*, the second either *q* or *r*. For each possible combination of the two feature variables, we have one example with the *x* outcome and one with the *y* outcome. Now suppose that our given context is *pq*. We get four possible supracontexts for this given context, of which only *pq* is homogeneous:

| supracontext | data items | disagreements | subcontextual partition | data items | disagreements |
|---|---|---|---|---|---|
| p q | pq**x** pq**y** | 2 | p q | pq**x** pq**y** | 2 |
| p - | pq**x** pq**y** pr**x** pr**y** | 8 | p q<br>p q̄ | pq**x** pq**y**<br>pr**x** pr**y** | 2<br>2 |
| - q | oq**x** oq**y** pq**x** pq**y** | 8 | p q<br>p̄ q | pq**x** pq**y**<br>oq**x** oq**y** | 2<br>2 |

| | | | | | | |
|---|---|---|---|---|---|---|
| - - | oq**x** | 32 | p  q | pq**x** pq**y** | 2 |
| | oq**y** | | p  q̄ | pr**x** pr**y** | 2 |
| | or**x** | | p̄  q | oq**x** oq**y** | 2 |
| | or**y** | | p̄  q̄ | or**x** or**y** | 2 |
| | pq**x** | | | | |
| | pq**y** | | | | |
| | pr**x** | | | | |
| | pr**y** | | | | |

In other words, in order to predict the behavior of *pq*, we are forced to rely solely upon the supracontext *pq*, which predicts a probability of 1/2 each for *x* and *y*. Yet for the three other supracontexts the relative frequencies of the *x* and *y* outcomes is an identical (1/2, 1/2). For further discussion of this issue, see section 1.9.

On the other hand, if the data set contained only instances of *pqx* and *pqy* (in the simplest case, only one instance of each), then all four supracontexts would be homogeneous:

| *supracontext* | *data items* | *disagreements* | *subcontextual partition* | *data items* | *disagreements* |
|---|---|---|---|---|---|
| p  q | pq**x** pq**y** | 2 | p  q | pq**x** pq**y** | 2 |
| p  - | pq**x** pq**y** | 2 | p  q<br>p  q̄ | pq**x** pq**y**<br>– | 2<br>0 |
| -  q | pq**x** pq**y** | 2 | p  q<br>p̄  q | pq**x** pq**y**<br>– | 2<br>0 |
| -  - | pq**x** pq**y** | 2 | p  q<br>p  q̄<br>p̄  q<br>p̄  q̄ | pq**x** pq**y**<br>–<br>–<br>– | 2<br>0<br>0<br>0 |

In the data set containing only these two instances, *pqx* and *pqy*, the feature variables *p* and *q* are wholly redundant in predicting *x* and *y*. Thus either feature variable (or even both) can be removed and the resulting supracontext remains homogeneous. What this all means is that a homogeneous nondeterministic supracontext can be generalized only for redundant feature variables (or combinations of redundant feature variables).

Another way to look at homogeneity is to consider it in terms of a two-way statistical array. In the following examples, the asterisks in the array show where the data occurrences are found. If they are all in one column, then the supracontextual array is deterministic in outcomes but generally nondeterministic in subcontexts:

DETERMINISTIC IN OUTCOMES

*outcomes*

|   |   | 1 | 2 | 3 | 4 | 5 | 6 |
|---|---|---|---|---|---|---|---|
|              | a | 0 | 0 | * | 0 | 0 | 0 |
|              | b | 0 | 0 | * | 0 | 0 | 0 |
| *subcontexts* | c | 0 | 0 | 0 | 0 | 0 | 0 |
|              | d | 0 | 0 | * | 0 | 0 | 0 |
|              | e | 0 | 0 | * | 0 | 0 | 0 |

If the asterisks are all in one row, then the supracontextual array is deterministic in subcontexts but generally nondeterministic in outcomes:

DETERMINISTIC IN SUBCONTEXTS

*outcomes*

|   |   | 1 | 2 | 3 | 4 | 5 | 6 |
|---|---|---|---|---|---|---|---|
|              | a | 0 | 0 | 0 | 0 | 0 | 0 |
|              | b | * | * | 0 | * | 0 | * |
| *subcontexts* | c | 0 | 0 | 0 | 0 | 0 | 0 |
|              | d | 0 | 0 | 0 | 0 | 0 | 0 |
|              | e | 0 | 0 | 0 | 0 | 0 | 0 |

The supracontext may also be deterministic in both outcomes and subcontexts. This is referred to as a singleton array; there can be more than one data item, but they are all identical with respect to the given context, having the same outcome and the same subcontext:

DETERMINISTIC IN OUTCOMES AND SUBCONTEXTS

*outcomes*

|   |   | 1 | 2 | 3 | 4 | 5 | 6 |
|---|---|---|---|---|---|---|---|
|              | a | 0 | 0 | 0 | 0 | 0 | 0 |
|              | b | 0 | 0 | * | 0 | 0 | 0 |
| *subcontexts* | c | 0 | 0 | 0 | 0 | 0 | 0 |
|              | d | 0 | 0 | 0 | 0 | 0 | 0 |
|              | e | 0 | 0 | 0 | 0 | 0 | 0 |

Of course, the supracontext can be empty, which gives us a null array:

EMPTY  SUPRACONTEXT

*outcomes*

1 2 3 4 5 6

|  |  |  |
|---|---|---|
|  | a | 0 0 0 0 0 0 |
|  | b | 0 0 0 0 0 0 |
| *subcontexts* | c | 0 0 0 0 0 0 |
|  | d | 0 0 0 0 0 0 |
|  | e | 0 0 0 0 0 0 |

On the other hand, heterogeneity occurs wherever the asterisks are found in more than one outcome and in more than one subcontext, as in the following two examples:

HETEROGENEOUS  SUPRACONTEXTS

*outcomes*                                                        *outcomes*

1 2 3 4 5 6                                                        1 2 3 4 5 6

|  |  |  |  |  |  |
|---|---|---|---|---|---|
|  | a | 0 0 0 0 0 0 |  | a | 0 0 0 0 0 0 |
|  | b | 0 * 0 0 0 0 |  | b | 0 * 0 0 * 0 |
| *subcontexts* | c | 0 0 0 0 0 0 | *subcontexts* | c | 0 0 0 0 0 * |
|  | d | 0 0 0 0 * 0 |  | d | 0 0 * 0 0 0 |
|  | e | 0 0 0 0 0 0 |  | e | * 0 0 0 0 * |

Finding a supracontext whose data items occur in more than one subcontext and exhibit more than one outcome is therefore equivalent to finding a supracontext whose subcontexts have less disagreements – in other words, the subcontexts behave differently (or potentially differently) than the supracontext.

1.8  *Statistical Power*

The determination of heterogeneity is the most powerful statistical test possible. If there is any evidence that a supracontext might be heterogeneous, Analogical Modeling will assume that it is. That is, if there is any sign that a subcontext might behave differently than its supracontext, then AM assumes that it is different.

 There is, nonetheless, a way to reduce the statistical power of AM, but without rejecting the very simple decision procedure of AM. This is accomplished by introducing imperfect memory, which in its simplest case assigns a probability of 1/2 that any given data item will be remembered. Behavior that is statistically significant will still hold even when about half the items in the data set are remembered. Using imperfect memory, we find that AM behaves as if it were following the usual restrictions on the power of a statistical test. For instance, under perfect memory, the estimated probability of occurrence for an outcome is always equal to the relative frequency of that outcome in the data set. But under imperfect memory set at 1/2, the estimated probability of occurrence will vary (depending on which data items are remembered), yet the variance for that estimated probability will quickly approach the standard unbiased estimate of variance as the number of data items for that given context increases. Another important statistical property of AM (again under the assumption of imperfect memory set at 1/2) is that the notion of level of significance can be defined in terms of the number of remembered items in the data set. For these properties and others, see Skousen 1998.

Analogical Modeling allows variability in its predictions, but only indirectly (in terms of imperfect memory). The same fundamental decision procedure is always used (namely, never allow a supracontext whose subcontextual partition reduces the number of disagreements). The normally expected statistical behavior is then derived indirectly, given that the chances of remembering a particular data item equals 1/2.

1.9 *The Rule Equivalent to Analogical Modeling*

Analogical Modeling can be reinterpreted in terms of rules, as follows: (1) every true rule exists; and (2) the probability of using a true rule is proportional to its frequency squared. A rule is defined in terms of a context of applicability and assigns a probability to each of the possible outcomes for that context. Basically, every occurring homogeneous supracontext defines a rule context. In addition, the relative frequencies of outcomes for that supracontext are converted into probabilities. For our example of *oma*, there are four true rules (see the summary at the end of section 1.5):

| rule context | probability of outcomes | frequency of the rule | squared frequency of the rule |
|---|---|---|---|
| o  m  - | 1/2, 1/2 | 2 | 4 |
| -  m  a | 1, 0 | 1 | 1 |
| o  -  - | 1/2, 1/2 | 2 | 4 |
| -  -  a | 1, 0 | 2 | 4 |

To get the probability of applying the rule, the chances are normed by the sum of the chances (here, $4 + 1 + 4 + 4 = 13$):

| rule context | probability of outcomes | frequency of the rule | probability of the rule |
|---|---|---|---|
| o  m  - | 1/2, 1/2 | 2 | 4/13 |
| -  m  a | 1, 0 | 1 | 1/13 |
| o  -  - | 1/2, 1/2 | 2 | 4/13 |
| -  -  a | 1, 0 | 2 | 4/13 |

We first select a rule according to its probability of application, then predict the outcome according to the probability of outcomes for that rule. This gives us the following results for the given context *oma*:

*probability of outcome* x:   $4/13 \times 1/2 \;+\; 1/13 \times 1 \;+\; 4/13 \times 1/2 \;+\; 4/13 \times 1 \;=\; 9/13$

*probability of outcome* y:   $4/13 \times 1/2 \;+\; 4/13 \times 1/2 \;=\; 4/13$

In this particular approach, one must first choose the rule, then choose the outcome for the chosen rule. The two-stage step of measurement found in quantum mechanics is clearly required when we view AM as a system of all possible true rules.

In order to determine whether a rule is a true one, we compare the behavior of the rule with its subrules. The subrules are the rule equivalents of the subcontexts defined by supracontexts. In particular, we have two properties equivalent to the two properties listed for homogeneous supracontexts:

(1) a deterministic rule is always a true rule since all of its subrules are also deterministic;

(2) a nondeterministic rule is a true rule if every subrule of it has either a zero frequency (the subrule is a nonrule) or the same frequency as the rule itself.

The other possible rules are either nonrules or false rules. The nonrules are rules with zero frequency, which means that their probability functions are undefined. The false rules are always nondeterministic rules that have two or more occurring subrules of lesser but nonzero frequency.

Another way to view these results is to consider the relationship between a rule and its subrules. A deterministic rule is always homogeneous and any occurring subrule (which has a more specific contextual specification) will also be deterministic and therefore homogeneous. On the other hand, if the homogeneous rule is nondeterministic, the only occurring homogeneous subrules are ones that have the same frequency of occurrence as the rule itself, which means that the more specific context for the subrule will contain some redundancy.

The analogical system of "rules" is not like traditional rule approaches. Rule approaches typically try to partition the contextual space (that is, they cover the entire contextual space and without any overlap). Instead, in Analogical Modeling *all* the true rules are said to exist; there will be overlapping rules, redundant rules, and rules based on as little as one occurrence. (We could say that even the false rules – the heterogeneous ones – also exist, though their probability of being used is zero.)

Another difference between AM's "rules" and traditional rules is that AM rules are created "on the fly"; they are not stored somewhere, waiting to be used.

A third, and most crucial, difference deals with the question of what kind of nondeterministic rules are homogeneous and therefore true rules. Suppose our rule equivalent is set up to predict the outcome $x$ or $y$ under the conditions $p$ and $q$ for the data set of eight items *oqx, oqy, orx, ory, pqx, pqy, prx,* and *pry* (as described in section 1.7). The traditional rule equivalent states that predicting $x$ and $y$ is equally probable, not only for the given context $pq$, but also for *every* supracontext. We get the following listing of possible rules:

| *rule context* | *data items* | *probability of outcomes* x *and* y | *frequency of the rule* |
|---|---|---|---|
| p  q | pqx, pqy | 1/2, 1/2 | 2 |
| p  - | pqx, pqy, prx, pry | 1/2, 1/2 | 4 |
| -  q | oqx, oqy, pqx, pqy | 1/2, 1/2 | 4 |
| -  - | oqx, oqy, orx, ory, pqx, pqy, prx, pry | 1/2, 1/2 | 8 |

Under a traditional rule approach, all four rule contexts would be true rules since their assigned probabilities are identical, but under AM this occurs only when the frequency of these nondeterministic rule contexts is the same. In other words, under the AM interpretation, only the rule based on the context $pq$ would be homogeneous; the three others would be heterogeneous. To get the others to be homogeneous, the data set must contain instances of only *pq* – none of *oq, or,* and *pr*. (See the discussion in section 1.7 when this same problem is treated from the point of view of exemplars rather than rules.) The ultimate reason for this difference is that in AM there are no actual probabilities, only exemplars; we cannot ever claim that two nondeterministic rules have the same probability function unless they are identical in exemplars. Despite this conceptual difference, it is worth observing that for this example we still get the same prediction for the probability of outcomes, 1/2 for *x* and 1/2 for *y*.

Part 2: A General Quantum Computing Algorithm for Analogical Modeling

2.1  *Representations*

In this first section of part 2, I list the symbols needed for representing AM and QAM. With respect to AM, I generally use the following indexing system:

| | |
|---|---|
| $i, n$ | from $i = 1$ to $n$, where $n$ is the number of feature variables |
| $\iota, \omega$ | from $\iota = 1$ to $\omega$, where $\omega$ is the number of outcome variables |
| $j, m$ | from $j = 1$ to $m$, where $m$ is the number of data items |
| $k, \ell$ | from $k = 1$ to $\ell$, where $\ell$ is the number of bits needed for a variable specification |

In addition, $I$ is used to index the superposition of $n$ independent feature variables, from $I = 1$ to $2^n$. Indexing is used as a symbolic convention for representing the various bits and qubits. In other words, the indexes themselves ($i$, $\iota$, $j$, and $k$) are not actual variables; for instance, the statement "from $j = 1$ to $m$ do A" simply means 'do A to each of the $m$ data items'.

For QAM we have the following general notational system:

**X**    a qubit vector of length $L$
$X_i$    individual qubit representation, where $X_i$ is the $i$th qubit in the qubit vector **X**;
        thus $\mathbf{X} = X_1 X_2 X_3 \cdots X_L$

Qubit vectors are a sequence of qubits, with an assumed length of $L$ (which can be different for various vectors). Qubit vectors are given in bold, such as **X**. We use subscripts to stand as an index for the individual qubits in a qubit vector **X**. For instance, if **X** had $L = 8$ qubits, we can represent each qubit as a nonbold form of **X** with an added subscript: that is, this particular qubit vector **X** can be represented as $X_1 X_2 X_3 X_4 X_5 X_6 X_7 X_8$. Here each qubit $X_i$ can theoretically take on all possible quantum state values. But in QAM, qubit vectors will be restricted to sequences of zeros and ones, such as $\mathbf{X} = 01011001$. Each individual qubit can be referred to by means of a subscript, such as $X_1 = 0$ and $X_8 = 1$.

$\mathbf{X}^{[j]}$    the $j$th qubit vector **X**

Sometimes we will refer to different qubit vectors that are all of the same type **X**. In order to refer to such qubits individually, we will use superscripts in square brackets, such as $\mathbf{X}^{[j]}$ to stand for the $j$th qubit vector **X**.

**X′**   the final evolved state of a qubit vector **X**
$X'_i$   the final evolved state of the *i*th qubit in the qubit vector **X**

+   The qubit **X** will evolve towards a resulting state that will then be copied (usually only one of its qubits) to some other qubit Y. We typically refer to the resulting state of **X** as **X′**, an evolved qubit vector. When copying, we often copy only $X_L'$, the evolved state of the last (or *L*th) qubit in the qubit vector **X**. After the copying, **X′** is typically returned to its original qubit vector **X** for subsequent processing of other qubits; this is accomplished by applying the operators in reverse order.

**A ... T**   substantive qubit vectors
**U ... Z**   auxiliary qubit vectors

Capital letters near the end of the alphabet will be reserved for the auxiliary qubits that are used to derive various results from substantive qubits or from other derived qubits that typically remain unchanged during the particular evolution under consideration. Capital letters near the end of the Greek alphabet may also be used to represent auxiliary qubits (especially for ones involving outcomes).

**0**   a qubit vector of zeros
**1**   a qubit vector of ones
0   a zero qubit
1   a one qubit

We use bold **0** and **1** to refer to qubit vectors of all zeros and all ones, but nonbold 0 and 1 to refer to individual qubits of zero or one. The initial state of a qubit vector **X** is generally set at specific values. If all the qubits are zeros, we represent that vector as **X** = **0**; if all the qubits are ones, we have **X** = **1**.

$\mathbb{X}$   the superpositioned qubit vector $\mathbb{X}$
$\mathbb{X}_i$   the *i*th qubit in the superpositioned qubit vector $\mathbb{X}$
$\mathbb{X}^{[j]}$   the *j*th superpositioned qubit vector $\mathbb{X}$

When a qubit vector **X** is in superposition, we represent it as $\mathbb{X}$. Typographical outlining will be used to stand for the simultaneous multiple representation of superpositioning. We use a subscript *i* to represent the *i*th qubit in the superpositioned qubit vector $\mathbb{X}$ and the bracketed superscript *j* to stand for the *j*th superpositioned qubit vector $\mathbb{X}$.

$\mathbb{0}$   a superpositioned qubit vector of zeros
$\mathbb{1}$   a superpositioned qubit vector of ones

If superpositioning is involved, we may have a superpositioned qubit vector of all zeros ($\mathbb{0}$) or all ones ($\mathbb{1}$).

2.2  *Fundamental Operators*

Given this terminology, we can define the fundamental operators on qubits, qubit vectors, superpositioned qubits, and superpositioned qubit vectors:

| *for a qubit vector* | *for an individual qubit* |
|---|---|
| NOT(**X**) | NOT($X_i$) |
| CNOT(**A**, **X**) | CNOT($A_j$, $X_i$) |
| CCNOT(**A**, **B**, **X**) | CCNOT($A_j$, $B_k$, $X_i$) |

| *for a superpositioned qubit vector* | *for an individual qubit in a superpositioned qubit vector* |
|---|---|
| NOT($\mathbb{X}$) | NOT($\mathbb{X}_i$) |
| CNOT($\mathbb{A}$, $\mathbb{X}$) | CNOT($\mathbb{A}_j$, $\mathbb{X}_i$) |
| CCNOT($\mathbb{A}$, $\mathbb{B}$, $\mathbb{X}$) | CCNOT($\mathbb{A}_j$, $\mathbb{B}_k$, $\mathbb{X}_i$) |

NOT($X_i$) reverses the state of qubit $X_i$, CNOT($A_j$, $X_i$) reverses the state of qubit $X_i$ providing $A_j$ equals 1, and CCNOT($A_j$, $B_k$, $X_i$) reverses the state of $X_i$ providing both $A_j$ and $B_k$ equal 1. The capital letter C in CNOT and CCNOT stands for 'control'. In the same way, application of one of these operators to the qubit vector **X** means that every qubit $X_i$ in **X** is reversed (under the appropriate controls). Similarly, application to the superpositioned qubit vector $\mathbb{X}$ means that every qubit in every superposition of **X** is reversed (again under the appropriate controls). On the other hand, application to $\mathbb{X}_i$ means that the operator applies to the qubit $X_i$ (but only that qubit), yet in every superposition of $\mathbb{X}_i$.

In this paper, CCNOT will be considered a primitive operator, even though it can be derived from more fundamental operators working on single qubits and pairs of qubits (Barenco et al. 1995). Also note that the application of the operator to a qubit vector **X** such as NOT(**X**) can be theoretically considered the simultaneous or sequential application of the operator for each individual qubit $X_i$ in the vector **X**. In general, I will assume a simultaneous application.

By applying certain sequences of operators, we can create various derived operators. In general, we will use all caps to represent the name of such derived operators, just as we do with NOT, CNOT, and CCNOT. Within parentheses after the name of the operator we will specify the operands for the operator – namely, those (superpositioned) qubits and (superpositioned) qubit vectors that this operator will operate on. Thus we might have something like OPERATOR(**P**, **Q**, **Z**). By reversing the order of its fundamental operators, we can reverse the effect of the derived operator and return to the original states for the appropriate qubits. We will use the superscript -1 with the name of the derived operator to indicate its reversal, such as OPERATOR$^{-1}$(**P**, **Q**, **Z**).

### 2.3 *General Derived Operators*

Here I discuss a number of general operators derived from the more fundamental operators; these will be used throughout the rest of this paper. The same operators can also be applied to qubit vectors in superposition.

### 2.3.1 COMPARE(**A**, **B**, **X** = **1**)

COMPARE allows a qubit vector **X** of length *L* (where **X** is initially equal to all ones) to evolve such that it shows where two qubit vectors **A** and **B** each of length *L* differ. One should note that the auxiliary qubit vector **X** must be all ones prior to the operator applying. We represent this precondition for applying COMPARE as **X** = **1**, which must not be interpreted as setting **X** to all ones (as if **X** := **1**). The latter could be misinterpreted as specifying that **X** must be initially set to all ones as part of the operation. In general, such a resetting of states would be irreversible and thus unallowable. Instead, **X** must start out equal to all ones prior to using COMPARE the first time. If COMPARE is used later on, we must then make sure that **X** has already reverted to all ones before reapplying COMPARE. We cannot do this by simply resetting **X** to all ones; rather, we are required to reverse the auxiliary operators that were originally applied, thus getting back the original state (in this case) of all ones for **X**.

In the following definition of COMPARE, the qubit vector **X** will evolve and will not, in general, remain as **X** = **1**. Thus the **X** in the second operator CNOT(**B**, **X**) is usually not identical to the original **X**, but is the **X** that results from first applying CNOT(**A**, **X**):

|  |  | 1 |  |  |  | $L = 5$ |
|---|---|---|---|---|---|---|
|  | **A** | 1 | 0 | 0 | 1 | 0 |
|  | **B** | 1 | 0 | 1 | 0 | 0 |
|  | **X** | 1 | 1 | 1 | 1 | 1 |
| CNOT(**A**, **X**) | **X** | 0 | 1 | 1 | 0 | 1 |
| CNOT(**B**, **X**) | **X**′ | 1 | 1 | 0 | 0 | 1 |

In other words, COMPARE(**A**, **B**, **X** = **1**) is equivalent to the following sequence of fundamental operators:

$$\text{COMPARE}(\mathbf{A}, \mathbf{B}, \mathbf{X} = \mathbf{1}) \equiv \text{CNOT}(\mathbf{A}, \mathbf{X} = \mathbf{1}) \, \text{CNOT}(\mathbf{B}, \mathbf{X})$$

In representing a sequence of operators, we apply the operators from left to right, although in this particular case the opposite sequence would give the same results. But in general, the sequence will make a difference (as seen below in the discussion regarding the operator ONES).

The qubit vector **X** evolves according to the specific vectors **A** and **B**, but **A** and **B** are left unchanged. We represent the final content of **X** for COMPARE(**A**, **B**, **X**) as **X**′ to show that the result may be different from the original **X**. But in general, **X** stands for the vector without respect to its particular values and at any stage in its evolution (including **X**′).

To reverse COMPARE(**A**, **B**, **X**), we reverse the sequence of the more fundamental operators:

$$\text{COMPARE}^{-1}(\mathbf{A}, \mathbf{B}, \mathbf{X}) \equiv \text{CNOT}(\mathbf{B}, \mathbf{X}) \, \text{CNOT}(\mathbf{A}, \mathbf{X})$$

|  |  | 1 |  |  |  | $L = 5$ |
|---|---|---|---|---|---|---|
|  | **A** | 1 | 0 | 0 | 1 | 0 |
|  | **B** | 1 | 0 | 1 | 0 | 0 |
|  | **X**′ | 1 | 1 | 0 | 0 | 1 |
| CNOT(**B**, **X**) | **X** | 0 | 1 | 1 | 0 | 1 |
| CNOT(**A**, **X**) | **X** | 1 | 1 | 1 | 1 | 1 |

Thus we end up with our original state, **X** = **1**. In this particular case, we would get the same results if these two fundamental operators were repeated in their original order. But in general, we must reverse the order of the fundamental operators in order to get back to the original state for **X** starting from **X**′.

As an example, let us show how COMPARE determines where two letters such as *o* and *g* differ; in the following, we use the ASCII representation for the letters:

    COMPARE(**A** = *o*, **B** = *g*, **X** = **1**)

|  |  |  |  |
|---|---|---|---|
|  | *o* |  | 01101111 |
|  | *g* |  | 01100111 |
|  | **X** |  | 11111111 |
| CNOT(*o*, **X**) | **X** |  | 10010000 |
| CNOT(*g*, **X**) | **X**′ |  | 11110111 |

But when we compare *o* against itself, we end up with no change in **X**:

    COMPARE(**A** = *o*, **B** = *o*, **X** = **1**)

|  |  |  |  |
|---|---|---|---|
|  | *o* |  | 01101111 |
|  | *o* |  | 01101111 |
|  | **X** |  | 11111111 |
| CNOT(*o*, **X**) | **X** |  | 10010000 |
| CNOT(*o*, **X**) | **X**′ |  | 11111111 |

Thus the evolved state **X**′ indicates that there is no difference between *o* and *o*. In general, the original state **X** = **1** essentially assumes that the two items to be compared are identical, but if the qubit vectors **A** and **B** differ, then the particular feature variables that differ will show up as zeros at the end of the evolution of **X** to **X**′.

2.3.2  ONES(**X**, **Y** = **0**)

The operator ONES works on a qubit vector **Y** of length $L+1$; initially all $L+1$ qubits are equal to zeros (that is, $Y_0 = 0$, $Y_1 = 0$, $Y_2 = 0$, $\cdots$, $Y_L = 0$). The initial qubit $Y_0$ is negated to one and then we work through the qubit vector **X** looking for zeros. The qubit vector **Y** will evolve such that ultimately it will tell us whether a qubit vector **X** of length $L$ is composed of all ones or not. When **Y**′ is obtained, the last qubit of **Y** holds the answer. In other words, $Y'_L$ tells us whether a qubit vector **X** has only ones: if $Y'_L$ equals zero, then **X** contains one or more zeros; if $Y'_L$ equals one, then **X** contains only ones.

The sequence of fundamental operators for ONES(**X**, **Y** = **0**) is as follows:

    NOT($Y_0$)
    from $k = 1$ to $L$ do
        CCNOT($X_k$, $Y_{k-1}$, $Y_k$)

That is, we move across **X** looking for zeros. If we find at least one, $Y_L$ will end up as zero; otherwise, $Y_L$ will end up as one. We provide an example for each of these cases:

(a) There is at least one zero in the vector **X**:

|   |   |   | 0 | 1 | ... |   |   | L = 5 |
|---|---|---|---|---|---|---|---|---|
|   |   | **X** | 1 | 1 | 0 | 0 | 1 |   |
|   |   | **Y** | 0 | 0 | 0 | 0 | 0 | 0 |
| k |   |   |   |   |   |   |   |   |
| 0 | NOT($Y_0$) | **Y** | 1 | 0 | 0 | 0 | 0 | 0 |
| 1 | CCNOT($X_k$, $Y_{k-1}$, $Y_k$) | **Y** | 1 | <u>1</u> | 0 | 0 | 0 | 0 |
| 2 | CCNOT($X_k$, $Y_{k-1}$, $Y_k$) | **Y** | 1 | 1 | <u>1</u> | 0 | 0 | 0 |
| 3 | CCNOT($X_k$, $Y_{k-1}$, $Y_k$) | **Y** | 1 | 1 | 1 | <u>0</u> | 0 | 0 |
| 4 | CCNOT($X_k$, $Y_{k-1}$, $Y_k$) | **Y** | 1 | 1 | 1 | 0 | <u>0</u> | 0 |
| 5 | CCNOT($X_k$, $Y_{k-1}$, $Y_k$) | **Y'** | 1 | 1 | 1 | 0 | 0 | <u>0</u> |

$$Y'_L = 0$$

(b) There are no zeros in the vector **X**:

|   |   |   | 0 | 1 | ... |   |   | L = 5 |
|---|---|---|---|---|---|---|---|---|
|   |   | **X** | 1 | 1 | 1 | 1 | 1 |   |
|   |   | **Y** | 0 | 0 | 0 | 0 | 0 | 0 |
| k |   |   |   |   |   |   |   |   |
| 0 | NOT($Y_0$) | **Y** | 1 | 0 | 0 | 0 | 0 | 0 |
| 1 | CCNOT($X_k$, $Y_{k-1}$, $Y_k$) | **Y** | 1 | <u>1</u> | 0 | 0 | 0 | 0 |
| 2 | CCNOT($X_k$, $Y_{k-1}$, $Y_k$) | **Y** | 1 | 1 | <u>1</u> | 0 | 0 | 0 |
| 3 | CCNOT($X_k$, $Y_{k-1}$, $Y_k$) | **Y** | 1 | 1 | 1 | <u>1</u> | 0 | 0 |
| 4 | CCNOT($X_k$, $Y_{k-1}$, $Y_k$) | **Y** | 1 | 1 | 1 | 1 | <u>1</u> | 0 |
| 5 | CCNOT($X_k$, $Y_{k-1}$, $Y_k$) | **Y'** | 1 | 1 | 1 | 1 | 1 | <u>1</u> |

$$Y'_L = 1$$

It is important to note that we need ONES to be general enough so that it can operate on any given supracontext in the superposition. Therefore, we cannot stop as soon as we find the first zero; we must go through the entire vector **X** of length *L*.

In reversing ONES, the reversal of the order for each fundamental operation in ONES must be strictly observed. Thus in terms of fundamental operators, ONES$^{-1}$(**X**, **Y**) is as follows:

    from $k = L$ to 1 do
        CCNOT($X_k$, $Y_{k-1}$, $Y_k$)
    NOT($Y_0$)

When we apply ONES$^{-1}$(**X**, **Y**) after ONES(**X**, **Y** = **0**), we end up with **Y** = **0** and the original state for **X** (as is required). We consider here the two previous examples, the first with at least one zero and the second with only ones:

(a) There is at least one zero in the vector **X** ($Y'_L = 0$):

|   |   |   |   | 0 | 1 |   | ... |   | L = 5 |
|---|---|---|---|---|---|---|-----|---|-------|
|   |   |   | **X** |   | 1 | 1 | 0 | 0 | 1 |
|   |   |   | **Y'** | 1 | 1 | 1 | 0 | 0 | 0 |
| k |   |   |   |   |   |   |   |   |   |
| 5 | CCNOT($X_k$, $Y_{k-1}$, $Y_k$) | **Y** |   | 1 | 1 | 1 | 0 | 0 | <u>0</u> |
| 4 | CCNOT($X_k$, $Y_{k-1}$, $Y_k$) | **Y** |   | 1 | 1 | 1 | 0 | <u>0</u> | 0 |
| 3 | CCNOT($X_k$, $Y_{k-1}$, $Y_k$) | **Y** |   | 1 | 1 | 1 | <u>0</u> | 0 | 0 |
| 2 | CCNOT($X_k$, $Y_{k-1}$, $Y_k$) | **Y** |   | 1 | 1 | <u>0</u> | 0 | 0 | 0 |
| 1 | CCNOT($X_k$, $Y_{k-1}$, $Y_k$) | **Y** |   | 1 | <u>0</u> | 0 | 0 | 0 | 0 |
| 0 | NOT($Y_0$) | **Y** |   | 0 | 0 | 0 | 0 | 0 | 0 |

(b) There are no zeros in the vector **X** ($Y'_L = 1$):

|   |   |   |   | 0 | 1 |   | ... |   | L = 5 |
|---|---|---|---|---|---|---|-----|---|-------|
|   |   |   | **X** |   | 1 | 1 | 1 | 1 | 1 |
|   |   |   | **Y'** | 1 | 1 | 1 | 1 | 1 | 1 |
| k |   |   |   |   |   |   |   |   |   |
| 5 | CCNOT($X_k$, $Y_{k-1}$, $Y_k$) | **Y** |   | 1 | 1 | 1 | 1 | 1 | <u>0</u> |
| 4 | CCNOT($X_k$, $Y_{k-1}$, $Y_k$) | **Y** |   | 1 | 1 | 1 | 1 | <u>0</u> | 0 |
| 3 | CCNOT($X_k$, $Y_{k-1}$, $Y_k$) | **Y** |   | 1 | 1 | 1 | <u>0</u> | 0 | 0 |
| 2 | CCNOT($X_k$, $Y_{k-1}$, $Y_k$) | **Y** |   | 1 | 1 | <u>0</u> | 0 | 0 | 0 |
| 1 | CCNOT($X_k$, $Y_{k-1}$, $Y_k$) | **Y** |   | 1 | <u>0</u> | 0 | 0 | 0 | 0 |
| 0 | NOT($Y_0$) | **Y** |   | <u>0</u> | 0 | 0 | 0 | 0 | 0 |

But just repeating the sequence in the same order rather the than reverse order would not, in general, lead to restoring the original state. In the following, **Y** starts out as **Y'** = 111000, the result of previously applying ONES(**X**, **Y** = **0**) to **X** = 11001; if we follow the exact same sequence for ONES$^{-1}$ as for ONES, we do not restore **Y** = **0** but rather **Y** = 010000:

|   |   |   |   | 0 | 1 |   | ... |   | L = 5 |
|---|---|---|---|---|---|---|-----|---|-------|
|   |   |   | **X** |   | 1 | 1 | 0 | 0 | 1 |
|   |   |   | **Y'** | 1 | 1 | 1 | 0 | 0 | 0 |
| k |   |   |   |   |   |   |   |   |   |
| 0 | NOT($Y_0$) | **Y** |   | <u>0</u> | 1 | 1 | 0 | 0 | 0 |
| 1 | CCNOT($X_k$, $Y_{k-1}$, $Y_k$) | **Y** |   | 0 | <u>1</u> | 1 | 0 | 0 | 0 |
| 2 | CCNOT($X_k$, $Y_{k-1}$, $Y_k$) | **Y** |   | 0 | 1 | <u>0</u> | 0 | 0 | 0 |
| 3 | CCNOT($X_k$, $Y_{k-1}$, $Y_k$) | **Y** |   | 0 | 1 | 0 | <u>0</u> | 0 | 0 |
| 4 | CCNOT($X_k$, $Y_{k-1}$, $Y_k$) | **Y** |   | 0 | 1 | 0 | 0 | <u>0</u> | 0 |
| 5 | CCNOT($X_k$, $Y_{k-1}$, $Y_k$) | **Y** |   | 0 | 1 | 0 | 0 | 0 | <u>0</u> |

### 2.3.3 IDENTITY(**A**, **B**, **X** = **1**, **Y** = **0**)

Given two qubit vectors **A** and **B**, each of length *L*, we wish to determine whether **A** and **B** are identical. We first apply COMPARE(**A**, **B**, **X** = **1**), followed by ONES(**X**, **Y** = **0**). In other words, **X** first evolves to determine those qubits where **A** and **B** differ. For any qubit for which **A** and **B** differ, **X** will contain a zero. If **Y** ends up with a zero as its last qubit, then there is at least one differing qubit between **A** and **B**:

| | | | 0 | 1 | ... | | $L = 5$ |
|---|---|---|---|---|---|---|---|
| COMPARE(**A**, **B**, **X** = **1**) | | | | | | | |
| | | **A** | 1 | 0 | 0 | 1 | 0 |
| | | **B** | 1 | 0 | 1 | 0 | 0 |
| | | **X** | 1 | 1 | 1 | 1 | 1 |
| CNOT(**A**, **X**) | | **X** | 0 | 1 | 1 | 0 | 1 |
| CNOT(**B**, **X**) | | **X**′ | 1 | 1 | 0 | 0 | 1 |
| ONES(**X**, **Y** = **0**) | | | | | | | |
| | | **Y** | 0 | 0 | 0 | 0 | 0 | 0 |
| *k* | | | | | | | | |
| 0 | NOT($Y_0$) | **Y** | 1 | 0 | 0 | 0 | 0 | 0 |
| 1 | CCNOT($X_k$, $Y_{k-1}$, $Y_k$) | **Y** | 1 | <u>1</u> | 0 | 0 | 0 | 0 |
| 2 | CCNOT($X_k$, $Y_{k-1}$, $Y_k$) | **Y** | 1 | 1 | <u>1</u> | 0 | 0 | 0 |
| 3 | CCNOT($X_k$, $Y_{k-1}$, $Y_k$) | **Y** | 1 | 1 | 1 | <u>0</u> | 0 | 0 |
| 4 | CCNOT($X_k$, $Y_{k-1}$, $Y_k$) | **Y** | 1 | 1 | 1 | 0 | <u>0</u> | 0 |
| 5 | CCNOT($X_k$, $Y_{k-1}$, $Y_k$) | **Y** | 1 | 1 | 1 | 0 | 0 | <u>0</u> |

The result will found in the last qubit of vector **Y**. In this example, $Y'_L = 0$; in other words, **A** and **B** are different. However, if **A** and **B** are identical, then we will get $Y'_L = 1$. The reverse of IDENTITY is IDENTITY$^{-1}$(**A**, **B**, **X**, **Y**) ≡ ONES$^{-1}$(**X**, **Y**) COMPARE$^{-1}$(**A**, **B**, **X**), which means that **X** returns to **1** and **Y** to **0**.

### 2.4 *Preparing the Data for Analogical Modeling*

The initial key process in AM involves a comparison between each data item with the given context. Thus AM determines the difference vector **D** between two different items, say a data item *gfa*, which will be compared with a given context *oma* (as discussed earlier in part 1). For this example, the length of the item is *n* = 3, so we will have a qubit vector **D** of this length that will show which feature variables differ in comparing the two items. In order to compare *gfa* with *oma*, we compare *g* with *o*, *f* with *m*, and *a* with *a*. The results of our comparison will be stored in **D** as 110, where 0 means no difference and 1 means some difference. In other words, for this example of *gfa* versus *oma*, 110 means that *g* differs from *o*, *f* differs from *m*, but *a* is identical to *a*. The **D** vector starts out as all ones and evolves to show where the feature variables (letters) are different for the two items. To do this, we use a DIFFERENCE operator repeatedly (in this instance, *n* = 3 times). For purposes of efficiency, we will also reverse the individual operators so that we can reuse the same **X** and **Y** qubit vectors as auxiliary vectors.

The DIFFERENCE operator applies the IDENTITY operator to a single feature variable, stores the result of that operator, and then reverses the IDENTITY operator so that the difference for the next feature variable can be determined. More precisely, IDENTITY(**A**, **B**, **X = 1**, **Y = 0**) specifies in the last qubit of **Y'** whether the qubit vectors **A** and **B** are identical or not. Before reversing IDENTITY, we copy the negation of $Y'_L$ into $D_i$, where $1 \leq i \leq n$. Copying is possible here because all the qubits take only the values 0 and 1. In other words, this copying process does not violate the no-cloning property of quantum computing since the qubit state is restricted to either 0 or 1 (Nielsen and Chuang 2000:24-25).

Thus in comparing *oma* against *gfa*, we do DIFFERENCE($\mathbf{A}^{[i]}$, $\mathbf{B}^{[i]}$, **X = 1**, **Y = 0**, $D_i = 1$) from $i = 1$ to 3; $D_i$ is initially equal to 1, which means that initially for the entire qubit vector we have **D = 1**. The letters are here represented by their ASCII specification. In this instance, the auxiliary variable **Y** has the length $L = \ell + 1$, where $\ell = 8$, the size of the ASCII specification. In the following, $Y'_\ell$ stands for the last qubit in the evolved **Y'**:

$i = 1$ (comparing *o* against *g*)

|  |  |  |
|---|---|---|
|  | $\mathbf{A}^{[1]} = o$ | 01101111 |
|  | $\mathbf{B}^{[1]} = g$ | 01100111 |
|  | **X = 1** | 11111111 |
|  | **Y = 0** | 000000000 |
| IDENTITY($\mathbf{A}^{[1]}$, $\mathbf{B}^{[1]}$, **X**, **Y**) | **X'** | 11110111 |
|  | **Y'** | 11111000<u>0</u> |
| CNOT($Y'_\ell$, $D_1 = 1$) |  | $Y'_\ell = 0 \rightarrow D_1 = 1$ |
| IDENTITY$^{-1}$($\mathbf{A}^{[1]}$, $\mathbf{B}^{[1]}$, **X**, **Y**) | **X** | 11111111 |
|  | **Y** | 000000000 |

$i = 2$ (comparing *m* against *f*):

|  |  |  |
|---|---|---|
|  | $\mathbf{A}^{[2]} = m$ | 01101101 |
|  | $\mathbf{B}^{[2]} = f$ | 01100110 |
|  | **X = 1** | 11111111 |
|  | **Y = 0** | 000000000 |
| IDENTITY($\mathbf{A}^{[2]}$, $\mathbf{B}^{[2]}$, **X**, **Y**) | **X'** | 11110100 |
|  | **Y'** | 11111000<u>0</u> |
| CNOT($Y'_\ell$, $D_2 = 1$) |  | $Y'_\ell = 0 \rightarrow D_2 = 1$ |
| IDENTITY$^{-1}$($\mathbf{A}^{[2]}$, $\mathbf{B}^{[2]}$, **X**, **Y**) | **X** | 11111111 |
|  | **Y** | 000000000 |

$i = 3$ (comparing *a* against *a*):

|  |  |  |
|---|---|---|
|  | $\mathbf{A}^{[3]} = a$ | 01100001 |
|  | $\mathbf{B}^{[3]} = a$ | 01100001 |
|  | $\mathbf{X} = \mathbf{1}$ | 11111111 |
|  | $\mathbf{Y} = \mathbf{0}$ | 000000000 |
| IDENTITY($\mathbf{A}^{[3]}, \mathbf{B}^{[3]}, \mathbf{X}, \mathbf{Y}$) | $\mathbf{X}'$ | 11111111 |
|  | $\mathbf{Y}'$ | 11111111$\underline{1}$ |
| CNOT($\mathbf{Y}'_\ell, \mathbf{D}_3 = 1$) |  | $\mathbf{Y}'_\ell = 1 \rightarrow \mathbf{D}_3 = 0$ |
| IDENTITY$^{-1}$($\mathbf{A}^{[3]}, \mathbf{B}^{[3]}, \mathbf{X}, \mathbf{Y}$) | $\mathbf{X}$ | 11111111 |
|  | $\mathbf{Y}$ | 000000000 |

Thus $\mathbf{D}$ evolves from initial $\mathbf{D} = \mathbf{1}$ to give the difference vector $\mathbf{D}' = 110$ (in this example, the difference vector $\mathbf{D}$ for *oma* compared against *gfa*). In other words, the first two feature variables differ, the third one does not (that is, the third one is identical).

The DIFFERENCE operator can be used to go through the entire dataset of *m* items to create *m* different qubit vectors $\mathbf{D}$, each of length *n*. We end up with $\mathbf{D}^{[1]}, \mathbf{D}^{[2]}, \ldots, \mathbf{D}^{[m]}$. In our example from part 1, we have $m = 6$ data items, each with $n = 3$ feature variables, and each feature variable is represented by an 8-qubit vector (representing the appropriate ASCII character):

| *j* | 1 | 2 | 3 | 4 | 5 | 6 |
|---|---|---|---|---|---|---|
| given context | oma | oma | oma | oma | oma | oma |
| data item | oms | gfa | cms | cma | omn | gfr |
| difference $\mathbf{D}^{[j]}$ | 001 | 110 | 101 | 100 | 001 | 111 |

2.5 *Superpositions of Supracontexts and Other Qubit Vectors*

Given *n* independent feature variables in a given context, there will be $2^n$ supracontexts. In doing QAM, we first need to determine which data items are included in each supracontext. From those included data items, we then determine which supracontexts are homogeneous. In QAM, we treat all the supracontexts in a single simultaneously occurring superposition $\mathbb{S}$. Each supracontext $\mathbf{S}$ can be represented as a qubit vector of zeros and ones, where a zero means that the specific value of the feature variable is ignored in determining whether a data item is included in that supracontext while a one means that the specific value must hold.

For instance, if the given context has three independent and unstructured feature variables, there will be eight ($2^3$) supracontexts in $\mathbb{S}$: 111, 110, 101, 011, 100, 010, 001, and 000. For instance, if the given context is *oma* (see part 1), then a supracontext such as 101 means that we include in 101 all those data items for which the first feature variable is *o* and the third feature variable is *a* but we ignore whatever value the data item might have for the second feature variable.

The operators are written so that all the occurring qubit vectors in the superposition $\mathbb{S}$ (namely, the supracontexts) are treated simultaneously. Ultimately, however, only one of these supracontexts will be observed. The superpositioned qubit vector $\mathbb{S}$ will remain in superposition until measurement occurs. Each superpositioned qubit vector will be represented by means of typographical outlining. Once $\mathbb{S}$ is measured, the resulting qubit vector will now simply be a specific qubit vector **S** of zeros and ones. There are other superpositioned qubit vectors that will be necessary to the evolution of the system. Altogether, we have three types of qubit vectors that will be in superposition:

(1) *substantive qubit vectors*      *number of qubits in the vector*

| | | |
|---|---|---|
| $\mathbb{S}$ | supracontext | $n$ |
| $\mathbb{C}$ | containment | $m$ |
| $\mathbb{P}$ | nonplurality of feature variables | $n$ |
| $\mathbb{Q}$ | nonplurality of outcome variables | $\omega$ |
| $\mathbb{M}$ | plurality of subcontexts | 1 |
| $\mathbb{N}$ | plurality of outcomes | 1 |
| $\mathbb{H}$ | homogeneity | 1 |
| $\mathbb{A}$ | amplitude | $m$ |

The substantive vectors evolve from $\mathbb{C}$ through to $\mathbb{A}$ while $\mathbb{S}$ is maintained in superposition. Measurement or observation occurs when the system has evolved to $\mathbb{A}$. The substantive vectors are not reversed.

(2) *auxiliary qubit vectors*      *number of qubits in the vector*

| | |
|---|---|
| $\mathbb{U}$ | $n$ |
| $\mathbb{V}$ | $n$ |
| $\mathbb{W}$ | $n$ |
| $\mathbb{Z}$ | $n + 1$ |
| $\mathbb{X}$ | $m$ |
| $\mathbb{Y}$ | $m + 1$ |
| $\mathbb{\Phi}$ | $\omega$ |
| $\mathbb{\Psi}$ | $\omega$ |
| $\mathbb{\Upsilon}$ | $\omega + 1$ |

The auxiliary vectors can be (and often are) used more than once while the system evolves from $\mathbb{C}$ through to $\mathbb{A}$. In general, the auxiliary vectors are reversed so that they can be used over and over as the system in superposition evolves.

(3) *constant qubit vectors*      *number of qubits in the vector*

| | | |
|---|---|---|
| **D** | differences | $n$ |
| **F** | feature variables | $n$ |
| **Ω** | outcome variables | $\omega$ |

The constant vectors are data specifiers that are only used and never evolve in any way while the system is in superposition. In order to emphasize their constant nature, these vectors are represented in bold rather than typographical outlining.

As noted in the beginning of this part, $m$ is the number of data items in the data set, $n$ the number of feature variables, and $\omega$ the number of outcome variables. For our simple example from part 1, these values are $m = 6$, $n = 3$, and $\omega = 1$ (given that there are only two outcomes $x$ and $y$).

2.6 *Determining which Data Items are in the Supracontexts*

The first operator that we use on the superposition of supracontexts is called INCLUSION. Its purpose is to use the superpositioned auxiliary qubit vectors $\mathbb{W}$ and $\mathbb{Z}$ in a systematic way to determine the content of the superpositioned containment vector $\mathbb{C}$ for the superpositioned supracontext vector $\mathbb{S}$:

    INCLUSION($\mathbb{S}$, $\mathbf{D}$, $\mathbb{W} = \mathbb{1}$, $\mathbb{Z} = \mathbb{0}$)

    Here $\mathbb{S}$, $\mathbb{W}$, and $\mathbb{Z}$ are superpositioned qubit vectors. The qubit vector $\mathbf{D}$ is the difference qubit vector that has already been determined; $\mathbf{D}$ gives the comparison of a particular data item with the given context and shows which feature variables differ. Since for a given operation of INCLUSION the qubit vector $\mathbf{D}$ has a constant value no matter what the supracontext is, $\mathbf{D}$ is not explicitly represented as being in superposition, although it would be. $\mathbb{S}$ is of length $n$ (there are $n$ feature variables), as are $\mathbf{D}$ and $\mathbb{W}$, while $\mathbb{Z}$ is of length $n + 1$.

We will repeatedly use the operator INCLUSION to determine a superpositioned qubit vector $\mathbb{C}$. This vector $\mathbb{C}$ will be of length $m$ and will specify which of the $m$ data items are contained within the superpositioned supracontext $\mathbb{S}$. While the supracontexts are in superposition, the containment vector $\mathbb{C}$ must also be in superposition. Initially, $\mathbb{C}$ will be equal to all zeros: $\mathbb{C} = \mathbb{0}$.

If we need to refer to a particular qubit vector in a superpositioned qubit vector $\mathbb{X}$, we will symbolize it as a qubit vector $\mathbf{X}$ and use some kind of appropriate superscripted index. For instance, since the superpositioned qubit vector $\mathbb{S}$ is based on $n$ (the number of feature variables), the appropriate indexing feature variable $I$ takes on values from 1 to $2^n$. We will refer to a particular qubit vector that belongs to the superpositioned qubit vector $\mathbb{S}$ as $\mathbf{S}^{[I]}$. The inclusion symbol $\in$ may be used to show this: $\mathbf{S}^{[I]} \in \mathbb{S}$. Similarly, we can refer to one of the particular qubit vectors $\mathbf{C}$ in the superpositioned qubit vector $\mathbb{C}$ as $\mathbf{C}^{[I]} \in \mathbb{C}$, where $I$ takes on values from 1 to $2^n$.

The specific operation INCLUSION($\mathbb{S}$, $\mathbf{D}^{[j]}$, $\mathbb{W} = \mathbb{1}$, $\mathbb{Z} = \mathbb{0}$) is defined for a particular difference vector $\mathbf{D}^{[j]}$ for data item $j$ as follows:

    CCNOT($\mathbb{S}$, $\mathbf{D}^{[j]}$, $\mathbb{W} = \mathbb{1}$)
    ONES($\mathbb{W}$, $\mathbb{Z} = \mathbb{0}$)

This is equivalent to specifying that for each $\mathbf{S}^{[I]} \in \mathbb{S}$ (that is, from $I = 1$ to $2^n$) we simultaneously do the following for the data item $j$:

CCNOT($\mathbf{S}^{[l]}, \mathbf{D}^{[j]}, \mathbf{W}^{[l]} = \mathbf{1}$)
ONES($\mathbf{W}^{[l]}, \mathbf{Z}^{[l]} = \mathbf{0}$)

We apply this to each of the *m* data items as follows: for each data item *j*, we apply INCLUSION, then copy the result of INCLUSION into the superpositioned containment vector $\mathbb{C}$ at position *j* (the *j*th qubit in the superpositioned $\mathbb{C}$), and then reverse INCLUSION:

from *j* = 1 to *m* do
    INCLUSION($\mathbb{S}, \mathbf{D}^{[j]}, \mathbb{W} = \mathbb{1}, \mathbb{Z} = \mathbb{0}$)
    CNOT($\mathbb{Z}_n, \mathbb{C}_j = \mathbb{0}$)
    INCLUSION$^{-1}$($\mathbb{S}, \mathbf{D}^{[j]}, \mathbb{W}, \mathbb{Z}$)

CNOT($\mathbb{Z}_n, \mathbb{C}_j = \mathbb{0}$) takes the last qubit of $\mathbb{Z}$ (labeled as $\mathbb{Z}_n$) and copies it to $\mathbb{C}_j$ before reversing INCLUSION. Initially, $\mathbb{C}_j$ is set to $\mathbb{0}$ for all *j*.

We get the following results for our simple example by applying INCLUSION and copying its results into the containment vector $\mathbb{C}$ for the supracontext vector $\mathbb{S}$:

| supracontext $\mathbf{S}^{[l]} \in \mathbb{S}$ (of length $n = 3$) | containment $\mathbf{C}^{[l]} \in \mathbb{C}$ (of length $m = 6$) |
|---|---|
| 111 | 000000 |
| 110 | 100010 |
| 101 | 000000 |
| 011 | 000100 |
| 100 | 100010 |
| 010 | 101110 |
| 001 | 010100 |
| 000 | 111111 |

In the following, we can see this sequence of operators applying for each supracontext $\mathbf{S}^{[l]}$ in the superposition. Here the necessary reversal of INCLUSION is not specified; the reversed operator INCLUSION$^{-1}$ restores the original states for the auxiliary qubits $\mathbb{W}$ and $\mathbb{Z}$, thus allowing progression through the *m* data items:

|   |   | oms oma | gfa oma | cms oma | cma oma | omn oma | gfr oma |
|---|---|---|---|---|---|---|---|
| 111 | supracontext $\mathbf{S}^{[1]} \in \mathbb{S}$ | 111 | 111 | 111 | 111 | 111 | 111 |
|   | difference $\mathbf{D}^{[j]}$ | 001 | 110 | 101 | 100 | 001 | 111 |
|   | $\mathbb{W} = \mathbb{1}$ | 111 | 111 | 111 | 111 | 111 | 111 |
|   | CCNOT($\mathbb{S}, \mathbf{D}^{[j]}, \mathbb{W}$) | 110 | 001 | 010 | 011 | 110 | 000 |
|   | $\mathbb{Z} = \mathbb{0}$ | 0000 | 0000 | 0000 | 0000 | 0000 | 0000 |
|   | ONES($\mathbb{W}, \mathbb{Z}$) | 111<u>0</u> | 100<u>0</u> | 100<u>0</u> | 100<u>0</u> | 111<u>0</u> | 100<u>0</u> |
|   | containment $\mathbb{C}$ | 0 | 0 | 0 | 0 | 0 | 0 |

|     |                                          | om-<br>om- | gf-<br>om- | cm-<br>om- | cm-<br>om- | om-<br>om- | gf-<br>om- |
| --- | ---------------------------------------- | ---------- | ---------- | ---------- | ---------- | ---------- | ---------- |
| 110 | supracontext $\mathbf{S}^{[2]} \in \mathbb{S}$ | 110        | 110        | 110        | 110        | 110        | 110        |
|     | difference $\mathbf{D}^{[j]}$            | 001        | 110        | 101        | 100        | 001        | 111        |
|     | $\mathbb{W} = \mathbb{1}$                | 111        | 111        | 111        | 111        | 111        | 111        |
|     | CCNOT($\mathbb{S}, \mathbf{D}^{[j]}, \mathbb{W}$) | 111 | 001        | 011        | 011        | 111        | 001        |
|     | $\mathbb{Z} = \mathbb{0}$                | 0000       | 0000       | 0000       | 0000       | 0000       | 0000       |
|     | ONES($\mathbb{W}, \mathbb{Z}$)           | 111<u>1</u> | 100<u>0</u> | 100<u>0</u> | 100<u>0</u> | 111<u>1</u> | 100<u>0</u> |
|     | containment $\mathbb{C}$                 | 1          | 0          | 0          | 0          | 1          | 0          |

|     |                                          | o-s<br>o-a | g-a<br>o-a | c-s<br>o-a | c-a<br>o-a | o-n<br>o-a | g-r<br>o-a |
| --- | ---------------------------------------- | ---------- | ---------- | ---------- | ---------- | ---------- | ---------- |
| 101 | supracontext $\mathbf{S}^{[3]} \in \mathbb{S}$ | 101        | 101        | 101        | 101        | 101        | 101        |
|     | difference $\mathbf{D}^{[j]}$            | 001        | 110        | 101        | 100        | 001        | 111        |
|     | $\mathbb{W} = \mathbb{1}$                | 111        | 111        | 111        | 111        | 111        | 111        |
|     | CCNOT($\mathbb{S}, \mathbf{D}^{[j]}, \mathbb{W}$) | 110 | 011        | 010        | 011        | 110        | 010        |
|     | $\mathbb{Z} = \mathbb{0}$                | 0000       | 0000       | 0000       | 0000       | 0000       | 0000       |
|     | ONES($\mathbb{W}, \mathbb{Z}$)           | 111<u>0</u> | 100<u>0</u> | 100<u>0</u> | 100<u>0</u> | 111<u>0</u> | 100<u>0</u> |
|     | containment $\mathbb{C}$                 | 0          | 0          | 0          | 0          | 0          | 0          |

|     |                                          | -ms<br>-ma | -fa<br>-ma | -ms<br>-ma | -ma<br>-ma | -mn<br>-ma | -fr<br>-ma |
| --- | ---------------------------------------- | ---------- | ---------- | ---------- | ---------- | ---------- | ---------- |
| 011 | supracontext $\mathbf{S}^{[4]} \in \mathbb{S}$ | 011        | 011        | 011        | 011        | 011        | 011        |
|     | difference $\mathbf{D}^{[j]}$            | 001        | 110        | 101        | 100        | 001        | 111        |
|     | $\mathbb{W} = \mathbb{1}$                | 111        | 111        | 111        | 111        | 111        | 111        |
|     | CCNOT($\mathbb{S}, \mathbf{D}^{[j]}, \mathbb{W}$) | 110 | 101        | 110        | 111        | 110        | 100        |
|     | $\mathbb{Z} = \mathbb{0}$                | 0000       | 0000       | 0000       | 0000       | 0000       | 0000       |
|     | ONES($\mathbb{W}, \mathbb{Z}$)           | 111<u>0</u> | 110<u>0</u> | 111<u>0</u> | 111<u>1</u> | 111<u>0</u> | 110<u>0</u> |
|     | containment $\mathbb{C}$                 | 0          | 0          | 0          | 1          | 0          | 0          |

|     |                                          | o--<br>o-- | g--<br>o-- | c--<br>o-- | c--<br>o-- | o--<br>o-- | g--<br>o-- |
| --- | ---------------------------------------- | ---------- | ---------- | ---------- | ---------- | ---------- | ---------- |
| 100 | supracontext $\mathbf{S}^{[5]} \in \mathbb{S}$ | 100        | 100        | 100        | 100        | 100        | 100        |
|     | difference $\mathbf{D}^{[j]}$            | 001        | 110        | 101        | 100        | 001        | 111        |
|     | $\mathbb{W} = \mathbb{1}$                | 111        | 111        | 111        | 111        | 111        | 111        |
|     | CCNOT($\mathbb{S}, \mathbf{D}^{[j]}, \mathbb{W}$) | 111 | 011        | 011        | 011        | 111        | 011        |
|     | $\mathbb{Z} = \mathbb{0}$                | 0000       | 0000       | 0000       | 0000       | 0000       | 0000       |
|     | ONES($\mathbb{W}, \mathbb{Z}$)           | 111<u>1</u> | 100<u>0</u> | 100<u>0</u> | 100<u>0</u> | 111<u>1</u> | 100<u>0</u> |
|     | containment $\mathbb{C}$                 | 1          | 0          | 0          | 0          | 1          | 0          |

|     |                                      | -m- | -f- | -m- | -m- | -m- | -f- |
|     |                                      | -m- | -m- | -m- | -m- | -m- | -m- |
| 010 | supracontext $\mathbf{S}^{[6]} \in \mathbb{S}$ | 010 | 010 | 010 | 010 | 010 | 010 |
|     | difference $\mathbf{D}^{[j]}$        | 001 | 110 | 101 | 100 | 001 | 111 |
|     | $\mathbb{W} = \mathbb{1}$            | 111 | 111 | 111 | 111 | 111 | 111 |
|     | CCNOT($\mathbb{S}, \mathbf{D}^{[j]}, \mathbb{W}$) | 111 | 101 | 111 | 111 | 111 | 101 |
|     | $\mathbb{Z} = \mathbb{0}$            | 0000 | 0000 | 0000 | 0000 | 0000 | 0000 |
|     | ONES($\mathbb{W}, \mathbb{Z}$)       | 111<u>1</u> | 110<u>0</u> | 111<u>1</u> | 111<u>1</u> | 111<u>1</u> | 110<u>0</u> |
|     | containment $\mathbb{C}$             | 1 | 0 | 1 | 1 | 1 | 0 |

|     |                                      | --s | --a | --s | --a | --n | --r |
|     |                                      | --a | --a | --a | --a | --a | --a |
| 001 | supracontext $\mathbf{S}^{[7]} \in \mathbb{S}$ | 001 | 001 | 001 | 001 | 001 | 001 |
|     | difference $\mathbf{D}^{[j]}$        | 001 | 110 | 101 | 100 | 001 | 111 |
|     | $\mathbb{W} = \mathbb{1}$            | 111 | 111 | 111 | 111 | 111 | 111 |
|     | CCNOT($\mathbb{S}, \mathbf{D}^{[j]}, \mathbb{W}$) | 110 | 111 | 110 | 111 | 110 | 110 |
|     | $\mathbb{Z} = \mathbb{0}$            | 0000 | 0000 | 0000 | 0000 | 0000 | 0000 |
|     | ONES($\mathbb{W}, \mathbb{Z}$)       | 111<u>0</u> | 111<u>1</u> | 111<u>0</u> | 111<u>1</u> | 111<u>0</u> | 111<u>0</u> |
|     | containment $\mathbb{C}$             | 0 | 1 | 0 | 1 | 0 | 0 |

|     |                                      | --- | --- | --- | --- | --- | --- |
|     |                                      | --- | --- | --- | --- | --- | --- |
| 000 | supracontext $\mathbf{S}^{[8]} \in \mathbb{S}$ | 000 | 000 | 000 | 000 | 000 | 000 |
|     | difference $\mathbf{D}^{[j]}$        | 001 | 110 | 101 | 100 | 001 | 111 |
|     | $\mathbb{W} = \mathbb{1}$            | 111 | 111 | 111 | 111 | 111 | 111 |
|     | CCNOT($\mathbb{S}, \mathbf{D}^{[j]}, \mathbb{W}$) | 111 | 111 | 111 | 111 | 111 | 111 |
|     | $\mathbb{Z} = \mathbb{0}$            | 0000 | 0000 | 0000 | 0000 | 0000 | 0000 |
|     | ONES($\mathbb{W}, \mathbb{Z}$)       | 111<u>1</u> | 111<u>1</u> | 111<u>1</u> | 111<u>1</u> | 111<u>1</u> | 111<u>1</u> |
|     | containment $\mathbb{C}$             | 1 | 1 | 1 | 1 | 1 | 1 |

2.7 *Determining the Plurality of Subcontexts and Outcomes for each Supracontext*

A supracontext is heterogenous whenever there is both a plurality of subcontexts and a plurality of outcomes within that supracontext. In determining heterogeneity, we consider only those data items that are contained within that supracontext. The superpositioned containment vector $\mathbb{C}$ specifies which data items are found in each supracontext. For instance, in our simple example, there are six data items; here I list the specific data items that are found in each supracontext:

| *supracontext* $\mathbf{S}^{[l]} \in \mathbb{S}$ | *containment* $\mathbf{C}^{[l]} \in \mathbb{C}$ | *fully specified contexts and outcomes* | | | | | |
|---|---|---|---|---|---|---|---|
| 111 | 000000 | — | — | — | — | — | — |
| 110 | 100010 | oms y | — | — | — | omn x | — |
| 101 | 000000 | — | — | — | — | — | — |
| 011 | 000100 | — | — | — | cma x | — | — |
| 100 | 100010 | oms y | — | — | — | omn x | — |
| 010 | 101110 | oms y | — | cms x | cma x | omn x | — |
| 001 | 010100 | — | gfa x | — | cma x | — | — |
| 000 | 111111 | oms y | gfa x | cms x | cma x | omn x | gfr x |

But the plurality of subcontexts is not the same thing as the plurality of the full data specification for a data item. As discussed under part 1, the subcontext specifies which contextual variables in a data item differ from the given context. In the above example, the given context is *oma*. The subcontextual specification for any given variable *z* in a data item will be the variable itself (*z*) if there is identity with the given context or the negation of that variable ($\bar{z}$) if there is nonidentity with the given context. Thus when we consider the subcontextual specification rather than the full contextual specification for each data item, we get the following for our simple example:

| *supracontext* $\mathbf{S}^{[l]} \in \mathbb{S}$ | *containment* $\mathbf{C}^{[l]} \in \mathbb{C}$ | *subcontexts and outcomes* | | | | | |
|---|---|---|---|---|---|---|---|
| 111 | 000000 | — | — | — | — | — | — |
| 110 | 100010 | omā y | — | — | — | omā x | — |
| 101 | 000000 | — | — | — | — | — | — |
| 011 | 000100 | — | — | — | ōma x | — | — |
| 100 | 100010 | omā y | — | — | — | omā x | — |
| 010 | 101110 | omā y | — | ōm̄a x | ōma x | omā x | — |
| 001 | 010100 | — | ōm̄a x | — | ōma x | — | — |
| 000 | 111111 | omā y | ōm̄a x | ōm̄a x | ōma x | omā x | ōm̄ā x |

For instance, in the supracontext 110, there are two occurring data items, *omsy* and *omnx*. We get a plurality of outcomes since there are two different outcomes that show up, *x* and *y*. But in the supracontext 110, the subcontexts for *oms* and *omn* state that the third feature variable is not *a*; thus as far as that supracontext is concerned, *oms* and *omn* are the same subcontextually (namely, *omā*). In other words, there is no plurality of subcontexts for the supracontext 110. As a consequence, supracontext 110 is homogeneous. To get heterogeneity, we must have plurality of outcomes and plurality of subcontexts.

In QAM we represent the subcontexts in terms of the difference vectors $\mathbf{D}^{[1]}, \mathbf{D}^{[2]}, \ldots, \mathbf{D}^{[m]}$, where each of *m* data items has its own difference vector $\mathbf{D}^{[j]}$. Where there is identity for a feature variable, we get a one; where there is nonidentity for a feature variable, we get a zero. Since we have only two outcomes, *x* and *y*, we can represent them respectively as one and zero (although one could also use their ASCII representations, which would mean that ω, the number of outcome variables, would be 8 instead of 1). This gives us the following for our simple example:

| supracontext $\mathbf{S}^{[I]} \in \mathbb{S}$ | containment $\mathbf{C}^{[I]} \in \mathbb{C}$ | difference vectors and binary outcomes | | | | | |
|---|---|---|---|---|---|---|---|
| 111 | 000000 | — | — | — | — | — | — |
| 110 | 100010 | 110 0 | — | — | — | 110 1 | — |
| 101 | 000000 | — | — | — | — | — | — |
| 011 | 000100 | — | — | — | 011 1 | — | — |
| 100 | 100010 | 110 0 | — | — | — | 110 1 | — |
| 010 | 101110 | 110 0 | — | 010 1 | 011 1 | 110 1 | — |
| 001 | 010100 | — | 001 1 | — | 011 1 | — | — |
| 000 | 111111 | 110 0 | 001 1 | 010 1 | 011 1 | 110 1 | 000 1 |

In order to determine heterogeneity for a given supracontext, two properties must be determined: (1) Is there a plurality of difference vectors for the data items contained within that supracontext? and (2) Is there a plurality of binary outcomes for the data items contained within that supracontext? For our simple example, we get the following results (with repetitions in italics):

| supracontext $\mathbf{S}^{[I]} \in \mathbb{S}$ | containment $\mathbf{C}^{[I]} \in \mathbb{C}$ | difference vectors | binary outcomes |
|---|---|---|---|
| 111 | 000000 | — | — |
| 110 | 100010 | 110, *110* | 0, 1 |
| 101 | 000000 | — | — |
| 011 | 000100 | 011 | 1 |
| 100 | 100010 | 110, *110* | 0, 1 |
| 010 | 101110 | 110, 010, 011, *110* | 0, 1, *1*, *1* |
| 001 | 010100 | 001, 011 | 1, *1* |
| 000 | 111111 | 110, 001, 010, 011, *110*, 000 | 0, 1, *1*, *1*, *1*, *1* |

By inspection, we can determine the plurality of the subcontexts (the difference vectors) and of the outcomes:

| supracontext $\mathbf{S}^{[I]} \in \mathbb{S}$ | containment $\mathbf{C}^{[I]} \in \mathbb{C}$ | plurality of subcontexts | plurality of outcomes |
|---|---|---|---|
| 111 | 000000 | no | no |
| 110 | 100010 | no | yes |
| 101 | 000000 | no | no |
| 011 | 000100 | no | no |
| 100 | 100010 | no | yes |
| 010 | 101110 | yes | yes |
| 001 | 010100 | yes | no |
| 000 | 111111 | yes | yes |

Heterogeneity occurs only when both the subcontexts and outcomes are plural (in this case, for the supracontexts 010 and 000).

Basically, the subcontexts for a given supracontext are nonplural if we never find any difference in a feature variable specification for the subcontexts in the supracontext. In other words, either a feature variable is always zero or it is always one for every data item found in the supracontext. Similarly, the outcomes for a given supracontext are nonplural if there is no evidence for any difference in the outcome for the data items in the supracontext (represented here as difference vectors).

To show how this is done, we begin by considering the subcontexts for all data items. In our simple example, we have six subcontexts, one for each data item:

    110    001    010    011    110    000

For each feature variable $i$, we determine the feature vector $\mathbf{F}^{[i]}$. These feature vectors are the subcontexts restricted to a single feature variable $i$:

| | | | | | | | |
|---|---|---|---|---|---|---|---|
| $\mathbf{F}^{[1]}$ | 1 | 0 | 0 | 0 | 1 | 0 | the first feature variable |
| $\mathbf{F}^{[2]}$ | 1 | 0 | 1 | 1 | 1 | 0 | the second feature variable |
| $\mathbf{F}^{[3]}$ | 0 | 1 | 0 | 1 | 0 | 0 | the third feature variable |

In other words, $\mathbf{F}^{[i]}$ represents the $i$th feature variable for all six data items. Similarly, we can represent the outcome for the data items in terms of its own binary variables. Since only one binary variable is needed to specify the two outcomes $x$ and $y$, there is only one outcome variable:

    $\mathbf{\Omega}^{[1]}$    0    1    1    1    1    1    the (only) outcome variable

Our method of hunting for plurality starts out by assuming nonplurality for both feature variables and outcome variables. We first test a feature or outcome vector to see if that vector has only zeros for the data items in the supracontext. We then negate each qubit in that feature or outcome vector and see if the inverted vector has only zeros for the data items in the supracontext. This second stage is equivalent to determining if the feature or outcome vector has only ones. If the vector has only zeros or only ones, then that feature or outcome variable is invariant. And if such invariance holds for all the feature variables, then the subcontext is nonplural. Similarly, if every outcome variable is invariant, then the outcome is nonplural.

For each supracontext $\mathbb{S}$, we use the containment vector $\mathbb{C}$ to figure out whether the feature vectors and the outcome vectors are plural. The process is as follows: We use a qubit vector $\mathbb{X}$ of length $m$ (where $m$ is the number of data items in the data set). The qubit $\mathbb{X}$ is initially equal to all ones and evolves to show whether a feature vector $\mathbf{F}^{[i]}$ has a zero for every data item contained in a given supracontext. We store the results in a superpositioned qubit vector $\mathbb{U}_i$. The operators are then reversed to bring back $\mathbb{X}$ so that it is once more equal to all ones. $\mathbf{F}^{[i]}$ is now inverted to $\sim\mathbf{F}^{[i]}$ and then $\mathbb{X}$ evolves to show whether the inverted feature vector, $\sim\mathbf{F}^{[i]}$, has a zero for every data item contained within the same supracontext (the tilde before $\mathbf{F}^{[i]}$ stands for the negation of each qubit in $\mathbf{F}^{[i]}$). Essentially, this means that we are determining whether $\mathbf{F}^{[i]}$ has a one for every data item contained within the same supracontext. We store these results for the inverted features in the superpositioned qubit vector $\mathbb{V}_i$. By combining the resulting states of $\mathbb{U}_i$ and $\mathbb{V}_i$ into the superpositioned state $\mathbb{P}_i$, we are able to determine whether that feature variable is the same for every data item in the supracontext. By doing this process for all $n$ feature variables, we are able to determine whether the subcontexts for a given supracontext are all nonplural. We do the same for the outcome variables; in our simple example, there is only one outcome variable, $\Omega^{[1]}$. More generally, however, we collect the results in $\Phi_i$ and $\Psi_i$, which are combined into the superpositioned $\mathbb{Q}_i$.

### 2.7.1 *Determining the Plurality of the Feature Variables*

We first show how to determine the plurality for the subcontexts in our simple example, then we turn to determining the plurality for the outcomes.

the first feature variable $\mathbf{F}^{[1]} = 100010$

$\mathbf{F}^{[1]}$ *is all zeros for every data item in the supracontext:*

| | | | | | | | | |
|---|---|---|---|---|---|---|---|---|
| supracontext $\mathbb{S}$ | 111 | 110 | 101 | 011 | 100 | 010 | 001 | 000 |
| containment $\mathbb{C}$ | 000000 | 100010 | 000000 | 000100 | 100010 | 101110 | 010100 | 111111 |
| feature $\mathbf{F}^{[1]}$ | 100010 | 100010 | 100010 | 100010 | 100010 | 100010 | 100010 | 100010 |
| $\mathbb{X} = \mathbb{1}$ | 111111 | 111111 | 111111 | 111111 | 111111 | 111111 | 111111 | 111111 |
| CCNOT($\mathbb{C}$, $\mathbf{F}^{[1]}$, $\mathbb{X}$) | 111111 | 011101 | 111111 | 111111 | 011101 | 011101 | 111111 | 011101 |
| $\mathbb{Y} = \mathbb{0}$ | 0000000 | 0000000 | 0000000 | 0000000 | 0000000 | 0000000 | 0000000 | 0000000 |
| ONES($\mathbb{X}$, $\mathbb{Y}$) | 111111<u>1</u> | 100000<u>0</u> | 111111<u>1</u> | 111111<u>1</u> | 100000<u>0</u> | 100000<u>0</u> | 111111<u>1</u> | 100000<u>0</u> |
| CNOT($\mathbb{Y}_m$, $\mathbb{U}_1 = \mathbb{0}$) | 1 | 0 | 1 | 1 | 0 | 0 | 1 | 0 |
| ONES$^{-1}$($\mathbb{X}$, $\mathbb{Y}$) | 0000000 | 0000000 | 0000000 | 0000000 | 0000000 | 0000000 | 0000000 | 0000000 |
| CCNOT($\mathbb{C}$, $\mathbf{F}^{[1]}$, $\mathbb{X}$) | 111111 | 111111 | 111111 | 111111 | 111111 | 111111 | 111111 | 111111 |

$\mathbf{F}^{[1]}$ *is all ones for every data item in the supracontext:*

| | | | | | | | | |
|---|---|---|---|---|---|---|---|---|
| supracontext $\mathbb{S}$ | 111 | 110 | 101 | 011 | 100 | 010 | 001 | 000 |
| containment $\mathbb{C}$ | 000000 | 100010 | 000000 | 000100 | 100010 | 101110 | 010100 | 111111 |
| feature $\mathbf{F}^{[1]}$ | 100010 | 100010 | 100010 | 100010 | 100010 | 100010 | 100010 | 100010 |
| NOT($\mathbf{F}^{[1]}$) ≡ ~$\mathbf{F}^{[1]}$ | 011101 | 011101 | 011101 | 011101 | 011101 | 011101 | 011101 | 011101 |
| $\mathbb{X} = \mathbb{1}$ | 111111 | 111111 | 111111 | 111111 | 111111 | 111111 | 111111 | 111111 |
| CCNOT($\mathbb{C}$, $\mathbf{F}^{[1]}$, $\mathbb{X}$) | 111111 | 111111 | 111111 | 111011 | 111111 | 110011 | 101011 | 100010 |
| $\mathbb{Y} = \mathbb{0}$ | 0000000 | 0000000 | 0000000 | 0000000 | 0000000 | 0000000 | 0000000 | 0000000 |
| ONES($\mathbb{X}$, $\mathbb{Y}$) | 111111<u>1</u> | 111111<u>1</u> | 111111<u>1</u> | 111100<u>0</u> | 111111<u>1</u> | 111000<u>0</u> | 110000<u>0</u> | 110000<u>0</u> |
| CNOT($\mathbb{Y}_m$, $\mathbb{V}_1 = \mathbb{0}$) | 1 | 1 | 1 | 0 | 1 | 0 | 0 | 0 |
| ONES$^{-1}$($\mathbb{X}$, $\mathbb{Y}$) | 0000000 | 0000000 | 0000000 | 0000000 | 0000000 | 0000000 | 0000000 | 0000000 |
| CCNOT($\mathbb{C}$, $\mathbf{F}^{[1]}$, $\mathbb{X}$) | 111111 | 111111 | 111111 | 111111 | 111111 | 111111 | 111111 | 111111 |
| NOT($\mathbf{F}^{[1]}$) | 100010 | 100010 | 100010 | 100010 | 100010 | 100010 | 100010 | 100010 |

summary for $\mathbf{F}^{[1]}$: *all zeros or all ones*

| supracontext $\mathbb{S}$ | 111 | 110 | 101 | 011 | 100 | 010 | 001 | 000 |
|---|---|---|---|---|---|---|---|---|
| $\mathbb{U}_1$ | 1 | 0 | 1 | 1 | 0 | 0 | 1 | 0 |
| $\mathbb{V}_1$ | 1 | 1 | 1 | 0 | 1 | 0 | 0 | 0 |
| NOT($\mathbb{U}_1$) | 0 | 1 | 0 | 0 | 1 | 1 | 0 | 1 |
| NOT($\mathbb{V}_1$) | 0 | 0 | 0 | 1 | 0 | 1 | 1 | 1 |
| CCNOT($\mathbb{U}_1$, $\mathbb{V}_1$, $\mathbb{P}_1 = \mathbb{1}$) | 1 | 1 | 1 | 1 | 1 | 0 | 1 | 0 |
| difference vectors | — | **1**10 | — | 0**1**1 | **1**10 | **1**10 | 00**1** | **1**10 |
|  |  | **1**10 |  | ↑ | **1**10 | 0**1**0 | 0**1**1 | 00**1** |
|  |  | ↑ |  |  | ↑ | 0**1**1 | ↑ | 0**1**0 |
|  |  |  |  |  |  | **1**10 |  | 0**1**1 |
|  |  |  |  |  |  | ↑ |  | **1**10 |
|  |  |  |  |  |  |  |  | **0**00 |
|  |  |  |  |  |  |  |  | ↑ |

the second feature variable $\mathbf{F}^{[2]} = 101110$

$\mathbf{F}^{[2]}$ *is all zeros for every data item in the supracontext:*

| supracontext $\mathbb{S}$ | 111 | 110 | 101 | 011 | 100 | 010 | 001 | 000 |
|---|---|---|---|---|---|---|---|---|
| containment $\mathbb{C}$ | 000000 | 100010 | 000000 | 000100 | 100010 | 101110 | 010100 | 111111 |
| feature $\mathbf{F}^{[2]}$ | 101110 | 101110 | 101110 | 101110 | 101110 | 101110 | 101110 | 101110 |
| $\mathbb{X} = \mathbb{1}$ | 111111 | 111111 | 111111 | 111111 | 111111 | 111111 | 111111 | 111111 |
| CCNOT($\mathbb{C}$, $\mathbf{F}^{[2]}$, $\mathbb{X}$) | 111111 | 011101 | 111111 | 111011 | 011101 | 010001 | 111011 | 010001 |
| $\mathbb{Y} = \mathbb{0}$ | 0000000 | 0000000 | 0000000 | 0000000 | 0000000 | 0000000 | 0000000 | 0000000 |
| ONES($\mathbb{X}$, $\mathbb{Y}$) | 111111<u>1</u> | 100000<u>0</u> | 111111<u>1</u> | 111100<u>0</u> | 100000<u>0</u> | 100000<u>0</u> | 111100<u>0</u> | 100000<u>0</u> |
| CNOT($\mathbb{Y}_m$, $\mathbb{U}_2 = \mathbb{0}$) | 1 | 0 | 1 | 0 | 0 | 0 | 0 | 0 |
| ONES$^{-1}$($\mathbb{X}$, $\mathbb{Y}$) | 0000000 | 0000000 | 0000000 | 0000000 | 0000000 | 0000000 | 0000000 | 0000000 |
| CCNOT($\mathbb{C}$, $\mathbf{F}^{[2]}$, $\mathbb{X}$) | 111111 | 111111 | 111111 | 111111 | 111111 | 111111 | 111111 | 111111 |

$\mathbf{F}^{[2]}$ *is all ones for every data item in the supracontext:*

| supracontext $\mathbb{S}$ | 111 | 110 | 101 | 011 | 100 | 010 | 001 | 000 |
|---|---|---|---|---|---|---|---|---|
| containment $\mathbb{C}$ | 000000 | 100010 | 000000 | 000100 | 100010 | 101110 | 010100 | 111111 |
| feature $\mathbf{F}^{[2]}$ | 101110 | 101110 | 101110 | 101110 | 101110 | 101110 | 101110 | 101110 |
| NOT($\mathbf{F}^{[2]}$) ≡ ~$\mathbf{F}^{[2]}$ | 010001 | 010001 | 010001 | 010001 | 010001 | 010001 | 010001 | 010001 |
| $\mathbb{X} = \mathbb{1}$ | 111111 | 111111 | 111111 | 111111 | 111111 | 111111 | 111111 | 111111 |
| CCNOT($\mathbb{C}$, $\mathbf{F}^{[2]}$, $\mathbb{X}$) | 111111 | 111111 | 111111 | 111111 | 111111 | 111111 | 101111 | 101110 |
| $\mathbb{Y} = \mathbb{0}$ | 0000000 | 0000000 | 0000000 | 0000000 | 0000000 | 0000000 | 0000000 | 0000000 |
| ONES($\mathbb{X}$, $\mathbb{Y}$) | 111111<u>1</u> | 111111<u>1</u> | 111111<u>1</u> | 111111<u>1</u> | 111111<u>1</u> | 111111<u>1</u> | 110000<u>0</u> | 110000<u>0</u> |
| CNOT($\mathbb{Y}_m$, $\mathbb{V}_2 = \mathbb{0}$) | 1 | 1 | 1 | 1 | 1 | 1 | 0 | 0 |
| ONES$^{-1}$($\mathbb{X}$, $\mathbb{Y}$) | 0000000 | 0000000 | 0000000 | 0000000 | 0000000 | 0000000 | 0000000 | 0000000 |
| CCNOT($\mathbb{C}$, $\mathbf{F}^{[2]}$, $\mathbb{X}$) | 111111 | 111111 | 111111 | 111111 | 111111 | 111111 | 111111 | 111111 |
| NOT($\mathbf{F}^{[2]}$) | 101110 | 101110 | 101110 | 101110 | 101110 | 101110 | 101110 | 101110 |

summary for **F**[2]: *all zeros or all ones*

| supracontext $\mathbb{S}$ | 111 | 110 | 101 | 011 | 100 | 010 | 001 | 000 |
|---|---|---|---|---|---|---|---|---|
| $\mathbb{U}_2$ | 1 | 0 | 1 | 0 | 0 | 0 | 0 | 0 |
| $\mathbb{V}_2$ | 1 | 1 | 1 | 1 | 1 | 1 | 0 | 0 |
| NOT($\mathbb{T}_2$) | 0 | 1 | 0 | 1 | 1 | 1 | 1 | 1 |
| NOT($\mathbb{V}_2$) | 0 | 0 | 0 | 0 | 0 | 0 | 1 | 1 |
| CCNOT($\mathbb{T}_2$, $\mathbb{V}_2$, $\mathbb{P}_2 = \mathbb{1}$) | 1 | 1 | 1 | 1 | 1 | 1 | 0 | 0 |
| difference vectors | — | 110 | — | 011 | 110 | 110 | 001 | 110 |
|  |  | 110 |  | ↑ | 110 | 010 | 011 | 001 |
|  |  | ↑ |  |  | ↑ | 011 | ↑ | 010 |
|  |  |  |  |  |  | 110 |  | 011 |
|  |  |  |  |  |  | ↑ |  | 110 |
|  |  |  |  |  |  |  |  | 000 |
|  |  |  |  |  |  |  |  | ↑ |

the third feature variable **F**[3] = 010100

**F**[3] *has all zeros for every data item in the supracontext:*

| supracontext $\mathbb{S}$ | 111 | 110 | 101 | 011 | 100 | 010 | 001 | 000 |
|---|---|---|---|---|---|---|---|---|
| containment $\mathbb{C}$ | 000000 | 100010 | 000000 | 000100 | 100010 | 101110 | 010100 | 111111 |
| feature **F**[3] | 010100 | 010100 | 010100 | 010100 | 010100 | 010100 | 010100 | 010100 |
| $\mathbb{X} = \mathbb{1}$ | 111111 | 111111 | 111111 | 111111 | 111111 | 111111 | 111111 | 111111 |
| CCNOT($\mathbb{C}$, **F**[3], $\mathbb{X}$) | 111111 | 111111 | 111111 | 111011 | 111111 | 111011 | 101011 | 101011 |
| $\mathbb{Y} = \mathbb{0}$ | 0000000 | 0000000 | 0000000 | 0000000 | 0000000 | 0000000 | 0000000 | 0000000 |
| ONES($\mathbb{X}$, $\mathbb{Y}$) | 1111111 | 1111111 | 1111111 | 1111000 | 1111111 | 1111000 | 1100000 | 1100000 |
| CNOT($\mathbb{Y}_m$, $\mathbb{U}_3 = \mathbb{0}$) | 1 | 1 | 1 | 0 | 1 | 0 | 0 | 0 |
| ONES$^{-1}$($\mathbb{X}$, $\mathbb{Y}$) | 0000000 | 0000000 | 0000000 | 0000000 | 0000000 | 0000000 | 0000000 | 0000000 |
| CCNOT($\mathbb{C}$, **F**[3], $\mathbb{X}$) | 111111 | 111111 | 111111 | 111111 | 111111 | 111111 | 111111 | 111111 |

**F**[3] *has all ones for every data item in the supracontext:*

| supracontext $\mathbb{S}$ | 111 | 110 | 101 | 011 | 100 | 010 | 001 | 000 |
|---|---|---|---|---|---|---|---|---|
| containment $\mathbb{C}$ | 000000 | 100010 | 000000 | 000100 | 100010 | 101110 | 010100 | 111111 |
| feature **F**[3] | 010100 | 010100 | 010100 | 010100 | 010100 | 010100 | 010100 | 010100 |
| NOT(**F**[3]) ≡ ~**F**[3] | 101011 | 101011 | 101011 | 101011 | 101011 | 101011 | 101011 | 101011 |
| $\mathbb{X} = \mathbb{1}$ | 111111 | 111111 | 111111 | 111111 | 111111 | 111111 | 111111 | 111111 |
| CCNOT($\mathbb{C}$, **F**[3], $\mathbb{X}$) | 111111 | 011101 | 111111 | 111111 | 011101 | 010101 | 111111 | 010100 |
| $\mathbb{Y} = \mathbb{0}$ | 0000000 | 0000000 | 0000000 | 0000000 | 0000000 | 0000000 | 0000000 | 0000000 |
| ONES($\mathbb{X}$, $\mathbb{Y}$) | 1111111 | 1000000 | 1111111 | 1111111 | 1000000 | 1000000 | 1111111 | 1000000 |
| CNOT($\mathbb{Y}_m$, $\mathbb{V}_3 = \mathbb{0}$) | 1 | 0 | 1 | 1 | 0 | 0 | 1 | 0 |
| ONES$^{-1}$($\mathbb{X}$, $\mathbb{Y}$) | 0000000 | 0000000 | 0000000 | 0000000 | 0000000 | 0000000 | 0000000 | 0000000 |
| CCNOT($\mathbb{C}$, **F**[3], $\mathbb{X}$) | 111111 | 111111 | 111111 | 111111 | 111111 | 111111 | 111111 | 111111 |
| NOT(**F**[3]) | 010100 | 010100 | 010100 | 010100 | 010100 | 010100 | 010100 | 010100 |

summary for $\mathbf{F}^{[3]}$: *all zeros or all ones*

| supracontext $\mathbb{S}$ | 111 | 110 | 101 | 011 | 100 | 010 | 001 | 000 |
|---|---|---|---|---|---|---|---|---|
| $\mathbb{U}_3$ | 1 | 1 | 1 | 0 | 1 | 0 | 0 | 0 |
| $\mathbb{V}_3$ | 1 | 0 | 1 | 1 | 0 | 0 | 1 | 0 |
| NOT($\mathbb{U}_3$) | 0 | 0 | 0 | 1 | 0 | 1 | 1 | 1 |
| NOT($\mathbb{V}_3$) | 0 | 1 | 0 | 0 | 1 | 1 | 0 | 1 |
| CCNOT($\mathbb{U}_3$, $\mathbb{V}_3$, $\mathbb{P}_3 = 1$) | 1 | 1 | 1 | 1 | 1 | 0 | 1 | 0 |
| | | | | | | | | |
| difference vectors | — | 11**0** | — | 01**1** | 11**0** | 11**0** | 00**1** | 11**0** |
| | | 11**0** | | ↑ | 11**0** | 01**0** | 01**1** | 00**1** |
| | | ↑ | | | ↑ | 01**1** | ↑ | 01**0** |
| | | | | | | 11**0** | | 01**1** |
| | | | | | | ↑ | | 11**0** |
| | | | | | | | | 00**0** |
| | | | | | | | | ↑ |

The superpositioned qubit vector $\mathbb{P}$ tells us whether each feature vector $\mathbf{F}^{[i]}$ is nonplural (a one means that it is nonplural). If any of the feature vectors is plural, then we say that the subcontexts for the supracontext are plural. This means that if $\mathbb{P}$ contains one or more zeros, then the subcontexts for the supracontext are plural. We use the operator ONES on $\mathbb{P}$ and assign its result to a superpositioned qubit $\mathbb{M}$, as follows:

> ONES($\mathbb{P}$, $\mathbb{Z} = \mathbb{0}$)
> CNOT($\mathbb{Z}_n$, $\mathbb{M} = 1$)
> ONES$^{-1}$($\mathbb{P}$, $\mathbb{Z}$)

So if $\mathbb{M}$ ends up with a zero for any given supracontext, then we have plurality for the subcontexts contained within that supracontext.

Applying this procedure to our simple example, we get the following results:

| supracontext $\mathbb{S}$ | 111 | 110 | 101 | 011 | 100 | 010 | 001 | 000 |
|---|---|---|---|---|---|---|---|---|
| $\mathbb{P}_1$ | 1 | 1 | 1 | 1 | 1 | 0 | 1 | 0 |
| $\mathbb{P}_2$ | 1 | 1 | 1 | 1 | 1 | 1 | 0 | 0 |
| $\mathbb{P}_3$ | 1 | 1 | 1 | 1 | 1 | 0 | 1 | 0 |
| | | | | | | | | |
| $\mathbb{P}$ | 111 | 111 | 111 | 111 | 111 | 010 | 101 | 000 |
| $\mathbb{Z} = \mathbb{0}$ | 0000 | 0000 | 0000 | 0000 | 0000 | 0000 | 0000 | 0000 |
| ONES($\mathbb{P}$, $\mathbb{Z}$) | 111**1** | 111**1** | 111**1** | 111**1** | 111**1** | 100**0** | 110**0** | 100**0** |
| CNOT($\mathbb{Z}_n$, $\mathbb{M} = 1$) | 0 | 0 | 0 | 0 | 0 | 1 | 1 | 1 |
| ONES$^{-1}$($\mathbb{P}$, $\mathbb{Z}$) | 0000 | 0000 | 0000 | 0000 | 0000 | 0000 | 0000 | 0000 |

In other words, the supracontexts 010, 001, and 000 each contain more than one subcontext:

| supracontext $\mathbb{S}$ | 111 | 110 | 101 | 011 | 100 | 010 | 001 | 000 |
|---|---|---|---|---|---|---|---|---|
| difference vectors | — | 110 | — | 011 | 110 | 110 | 001 | 110 |
| | | 110 | | | 110 | 010 | 011 | 001 |
| | | | | | | 011 | | 010 |
| | | | | | | 110 | | 011 |
| | | | | | | | | 110 |
| | | | | | | | | 000 |
| $\mathbb{M}$ (plurality of subcontexts) | 0 | 0 | 0 | 0 | 0 | 1 | 1 | 1 |

2.7.2 *Determining the Plurality of the Outcome Variables*

The plurality for the outcomes is determined similarly, with the results found in the single superpositioned qubit $\mathbb{N}$. In our simple example, there is only one outcome variable:

$\mathbf{\Omega}^{[1]}$   0  1  1  1  1  1

As before, we first test for all zeros and all ones, thus determining whether there is a plurality for outcomes for each supracontext in the superposition:

the (only) outcome variable $\mathbf{\Omega}^{[1]} = 011111$

$\mathbf{\Omega}^{[1]}$ *has all zeros for every data item in the supracontext:*

| supracontext $\mathbb{S}$ | 111 | 110 | 101 | 011 | 100 | 010 | 001 | 000 |
|---|---|---|---|---|---|---|---|---|
| containment $\mathbb{C}$ | 000000 | 100010 | 000000 | 000100 | 100010 | 101110 | 010100 | 111111 |
| feature $\mathbf{\Omega}^{[1]}$ | 011111 | 011111 | 011111 | 011111 | 011111 | 011111 | 011111 | 011111 |
| $\mathbb{X} = 1$ | 111111 | 111111 | 111111 | 111111 | 111111 | 111111 | 111111 | 111111 |
| CCNOT($\mathbb{C}$, $\mathbf{\Omega}^{[1]}$, $\mathbb{X}$) | 111111 | 111101 | 111111 | 111011 | 111101 | 110001 | 101011 | 100000 |
| $\mathbb{Y} = 0$ | 0000000 | 0000000 | 0000000 | 0000000 | 0000000 | 0000000 | 0000000 | 0000000 |
| ONES($\mathbb{X}$, $\mathbb{Y}$) | 111111<u>1</u> | 111110<u>0</u> | 111111<u>1</u> | 111100<u>0</u> | 111110<u>0</u> | 111000<u>0</u> | 110000<u>0</u> | 110000<u>0</u> |
| CNOT($\mathbb{Y}_m$, $\mathbb{\Phi}_1 = \mathbb{0}$) | 1 | 0 | 1 | 0 | 0 | 0 | 0 | 0 |
| ONES$^{-1}$($\mathbb{X}$, $\mathbb{Y}$) | 0000000 | 0000000 | 0000000 | 0000000 | 0000000 | 0000000 | 0000000 | 0000000 |
| CCNOT($\mathbb{C}$, $\mathbf{\Omega}^{[1]}$, $\mathbb{X}$) | 111111 | 111111 | 111111 | 111111 | 111111 | 111111 | 111111 | 111111 |

**Ω[1]** *has all ones for every data item in the supracontext:*

| supracontext $\mathbb{S}$ | 111 | 110 | 101 | 011 | 100 | 010 | 001 | 000 |
|---|---|---|---|---|---|---|---|---|
| containment $\mathbb{C}$ | 000000 | 100010 | 000000 | 000100 | 100010 | 101110 | 010100 | 111111 |
| feature $\mathbf{Ω}^{[1]}$ | 011111 | 011111 | 011111 | 011111 | 011111 | 011111 | 011111 | 011111 |
| NOT($\mathbf{Ω}^{[1]}$) ≡ ~$\mathbf{Ω}^{[1]}$ | 100000 | 100000 | 100000 | 100000 | 100000 | 100000 | 100000 | 100000 |
| $\mathbb{X} = \mathbb{1}$ | 111111 | 111111 | 111111 | 111111 | 111111 | 111111 | 111111 | 111111 |
| CCNOT($\mathbb{C}$, $\mathbf{Ω}^{[1]}$, $\mathbb{X}$) | 111111 | 011111 | 111111 | 111111 | 011111 | 011111 | 111111 | 011111 |
| $\mathbb{Y} = \mathbb{0}$ | 0000000 | 0000000 | 0000000 | 0000000 | 0000000 | 0000000 | 0000000 | 0000000 |
| ONES($\mathbb{X}$, $\mathbb{Y}$) | 1111111 | 1000000 | 1111111 | 1111111 | 1000000 | 1000000 | 1111111 | 1000000 |
| CNOT($\mathbb{Y}_m$, $\mathbb{Ψ}_1 = \mathbb{0}$) | 1 | 0 | 1 | 1 | 0 | 0 | 1 | 0 |
| ONES$^{-1}$($\mathbb{X}$, $\mathbb{Y}$) | 0000000 | 0000000 | 0000000 | 0000000 | 0000000 | 0000000 | 0000000 | 0000000 |
| CCNOT($\mathbb{C}$, $\mathbf{Ω}^{[1]}$, $\mathbb{X}$) | 111111 | 111111 | 111111 | 111111 | 111111 | 111111 | 111111 | 111111 |
| NOT($\mathbf{Ω}^{[1]}$) | 011111 | 011111 | 011111 | 011111 | 011111 | 011111 | 011111 | 011111 |

summary for $\mathbf{Ω}^{[1]}$: *all zeros or all ones*

| supracontext $\mathbb{S}$ | 111 | 110 | 101 | 011 | 100 | 010 | 001 | 000 |
|---|---|---|---|---|---|---|---|---|
| $\mathbb{Φ}_1$ | 1 | 0 | 1 | 0 | 0 | 0 | 0 | 0 |
| $\mathbb{Ψ}_1$ | 1 | 0 | 1 | 1 | 0 | 0 | 1 | 0 |
| NOT($\mathbb{Φ}_1$) | 0 | 1 | 0 | 1 | 1 | 1 | 1 | 1 |
| NOT($\mathbb{Ψ}_1$) | 0 | 1 | 0 | 0 | 1 | 1 | 0 | 1 |
| CCNOT($\mathbb{Φ}_1$, $\mathbb{Ψ}_1$, $\mathbb{Q}_1 = \mathbb{1}$) | 1 | 0 | 1 | 1 | 0 | 0 | 1 | 0 |

| binary outcomes | — | 0 | — | 1 | 0 | 0 | 1 | 0 |
|---|---|---|---|---|---|---|---|---|
| | | 1 | | ↑ | 1 | 1 | 1 | 1 |
| | | ↑ | | | ↑ | 1 | ↑ | 1 |
| | | | | | | 1 | | 1 |
| | | | | | | ↑ | | 1 |
| | | | | | | | | 1 |
| | | | | | | | | ↑ |

The superpositioned qubit vector $\mathbb{Q}$ tells us whether each feature vector $\mathbf{Ω}^{[i]}$ is nonplural (a one means that it is nonplural). In our simple example, however, there is only one outcome variable, $\mathbf{Ω}^{[1]}$, but we write our operators so they assume the more general approach that can handle any number of outcome variables. If any of the outcome vectors is plural, then we say that the outcomes for the supracontext are plural. This means that if $\mathbb{Q}$ contains one or more zeros, then the outcomes for the supracontext are plural. We will use the operator ONES on $\mathbb{Q}$ and assign its result to a superpositioned qubit $\mathbb{N}$, as follows:

> ONES($\mathbb{Q}$, $\mathbb{ϒ} = \mathbb{0}$)
> CNOT($\mathbb{ϒ}_ω$, $\mathbb{N} = \mathbb{1}$)
> ONES$^{-1}$($\mathbb{Q}$, $\mathbb{ϒ}$)

So if $\mathbb{N}$ ends up with a zero for any given supracontext, then we have plurality for the outcomes contained within that supracontext.

Applying this procedure to our simple example, we get the following results, noting that since there is only one outcome variable, $\mathbb{Q}_1$ is $\mathbb{Q}$:

| supracontext $\mathbb{S}$ | 111 | 110 | 101 | 011 | 100 | 010 | 001 | 000 |
|---|---|---|---|---|---|---|---|---|
| $\mathbb{Q}_1 \equiv \mathbb{Q}$ | 1 | 0 | 1 | 1 | 0 | 0 | 1 | 0 |
| $\Upsilon = \mathbb{0}$ | 00 | 00 | 00 | 00 | 00 | 00 | 00 | 00 |
| ONES($\mathbb{Q}, \Upsilon$) | 1<u>1</u> | 1<u>0</u> | 1<u>1</u> | 1<u>1</u> | 1<u>0</u> | 1<u>0</u> | 1<u>1</u> | 1<u>0</u> |
| CNOT($\Upsilon_\omega, \mathbb{N} = \mathbb{1}$) | 0 | 1 | 0 | 0 | 1 | 1 | 0 | 1 |
| ONES$^{-1}(\mathbb{Q}, \Upsilon$) | 00 | 00 | 00 | 00 | 00 | 00 | 00 | 00 |

In other words, the supracontexts 110, 100, 010, and 000 each contain more than one outcome (that is, these supracontexts are nondeterministic).

| supracontext $\mathbb{S}$ | 111 | 110 | 101 | 011 | 100 | 010 | 001 | 000 |
|---|---|---|---|---|---|---|---|---|
| binary outcomes | — | 0<br>1 | — | 1 | 0<br>1<br>1<br>1 | 0<br>1 | 1<br>1 | 0<br>1<br>1<br>1<br>1<br>1 |
| $\mathbb{N}$ (plurality of outcomes) | 0 | 1 | 0 | 0 | 1 | 1 | 0 | 1 |

2.7.3 *Combining the Two Properties of Plurality*

Finally, we combine the results of the single qubits $\mathbb{M}$ and $\mathbb{N}$ (that is, the plurality of the subcontexts and the plurality of the outcomes, but only for those data items contained within each superpositioned supracontext) in order to derive the homogeneity $\mathbb{H}$ for each supracontext. This is the desired property of our quantum evolution. A supracontext is heterogeneous (not homogeneous) when the supracontext has plurality of both subcontexts and outcomes; thus we have the following operator:

    CCNOT($\mathbb{M}, \mathbb{N}, \mathbb{H} = \mathbb{1}$)

For our simple example, we finally determine the desired property, the homogeneity $\mathbb{H}$:

| supracontext $\mathbb{S}$ | 111 | 110 | 101 | 011 | 100 | 010 | 001 | 000 |
|---|---|---|---|---|---|---|---|---|
| difference vectors with binary outcomes | — | 110 0<br>110 1 | — | 011 1 | 110 0<br>110 1 | 110 0<br>010 1<br>011 1<br>110 1 | 001 1<br>011 1 | 110 0<br>001 1<br>010 1<br>011 1<br>110 1<br>000 1 |
| $\mathbb{M}$ (plurality of subcontexts) | 0 | 0 | 0 | 0 | 0 | 1 | 1 | 1 |
| $\mathbb{N}$ (plurality of outcomes) | 0 | 1 | 0 | 0 | 1 | 1 | 0 | 1 |
| CCNOT($\mathbb{M}$, $\mathbb{N}$, $\mathbb{H} = \mathbb{1}$) | 1 | 1 | 1 | 1 | 1 | 0 | 1 | 0 |

Thus there are only two supracontexts that are heterogeneous, 010 and 000. All the rest are homogeneous.

2.8 *Observing the Results*

Just prior to observation of the quantum system, we identify those supracontexts that have this property of homogeneity and make their amplitude vector $\mathbb{A}$ equal to the containment vector $\mathbb{C}$. The superpositioned qubit vector $\mathbb{C}$ specifies which of the data items are contained in each supracontext. In our simple example, there are six data items; for each supracontext **S** we have the following containment vector **C**:

| supracontext $\mathbb{S}$ | 111 | 110 | 101 | 011 | 100 | 010 | 001 | 000 |
|---|---|---|---|---|---|---|---|---|
| containment $\mathbb{C}$ | 000000 | 100010 | 000000 | 000100 | 100010 | 101110 | 010100 | 111111 |

We now take $\mathbb{H}$ and for each supracontext in $\mathbb{C}$ determine the amplitude vector $\mathbb{A}$, as follows:

    from *j* = 1 to *m* do
        CCNOT($\mathbb{C}_j$, $\mathbb{H}$, $\mathbb{A}_j = \mathbb{0}$)

Basically, if H = 1 for a given supracontext, then **A** will be the same as **C** for that supracontext; but if H = 0 for the supracontext, then **A** will remain at **0**:

| supracontext $\mathbb{S}$ | 111 | 110 | 101 | 011 | 100 | 010 | 001 | 000 |
|---|---|---|---|---|---|---|---|---|
| containment $\mathbb{C}$ | 000000 | 100010 | 000000 | 000100 | 100010 | 101110 | 010100 | 111111 |
| homogeneity $\mathbb{H}$ | 1 | 1 | 1 | 1 | 1 | 0 | 1 | 0 |
| amplitude $\mathbb{A}$ | 000000 | 000000 | 000000 | 000000 | 000000 | 000000 | 000000 | 000000 |
| CCNOT($\mathbb{C}$, $\mathbb{H}$, $\mathbb{A}$) | 000000 | 100010 | 000000 | 000100 | 100010 | 000000 | 010100 | 000000 |
| data items with nonzero amplitude | — | 110 0<br>110 1 | — | 011 1 | 110 0<br>110 1 | — | 001 1<br>011 1 | — |

The amplitude represents the number of data items in each homogeneous supracontext; heterogeneous supracontexts have zero amplitude and can be said to be nonoccurring (or at least inaccessible). For each homogeneous supracontext, the amplitude is proportional to the frequency of occurrence for that supracontext.

In observing the superpositioned system, there are traditionally two stages of measurement. We first select one of the supracontexts. In other words, the superposition of supracontexts collapses to a single supracontext. Of course, only a homogeneous supracontext will be selected since the amplitude for each hetereogeneous supracontext is zero. The probability of selecting a particular supracontext is proportional to its amplitude squared – that is, if the supracontext is homogeneous, the probability is proportional to the square of that supracontext's frequency of occurrence, but the probability is zero if the supracontext is heterogeneous. The second stage of measurement is to randomly select one of the data items in that homogeneous supracontext, so that ultimately in any given measurement all we observe is the outcome for only one of the data items in that supracontext.

For our simple example, we get the following probabilities for selecting a particular supracontext:

| supracontext $\mathbb{S}$ | 111 | 110 | 101 | 011 | 100 | 010 | 001 | 000 |
|---|---|---|---|---|---|---|---|---|
| data items with nonzero amplitude | — | 110 0<br>110 1 | — | 011 1 | 110 0<br>110 1 | — | 001 1<br>011 1 | — |
| amplitude $\mathbb{A}$ | 000000 | 100010 | 000000 | 000100 | 100010 | 000000 | 010100 | 000000 |
| frequency | 0 | 2 | 0 | 1 | 2 | 0 | 2 | 0 |
| frequency squared | 0 | 4 | 0 | 1 | 4 | 0 | 4 | 0 |
| probability | 0 | 4/13 | 0 | 1/13 | 4/13 | 0 | 4/13 | 0 |

For a given supracontext, we have the following probabilities for selecting one of the two outcomes (0 or 1):

| supracontext $\mathbb{S}$ | 111 | 110 | 101 | 011 | 100 | 010 | 001 | 000 |
|---|---|---|---|---|---|---|---|---|
| data items with nonzero amplitude | — | 110 **0**<br>110 **1** | — | 011 **1** | 110 **0**<br>110 **1** | — | 001 **1**<br>011 **1** | — |
| outcome 0 | — | 1/2 | — | 0 | 1/2 | — | 0 | — |
| outcome 1 | — | 1/2 | — | 1 | 1/2 | — | 1 | — |

Combining the two stages, we get the overall probabilities for selecting a particular outcome (0 or 1) for the whole system:

| supracontext $\mathbb{S}$ | 111 | 110 | 101 | 011 | 100 | 010 | 001 | 000 |
|---|---|---|---|---|---|---|---|---|
| outcome 0 | — | 1/2 × 4/13 | — | 0 × 1/13 | 1/2 × 4/13 | — | 0 × 4/13 | — |
| outcome 1 | — | 1/2 × 4/13 | — | 1 × 1/13 | 1/2 × 4/13 | — | 1 × 4/13 | — |

Or by simplifying the fractions and explicitly stating that the probability of selecting a heterogeneous supracontext or a nonoccurring (technically homogeneous) supracontext is zero, we get the final results:

| supracontext $\mathbb{S}$ | 111 | 110 | 101 | 011 | 100 | 010 | 001 | 000 |
|---|---|---|---|---|---|---|---|---|
| outcome 0 | 0 | 2/13 | 0 | 0 | 2/13 | 0 | 0 | 0 |
| outcome 1 | 0 | 2/13 | 0 | 1/13 | 2/13 | 0 | 4/13 | 0 |

However, there is much simpler way to observe (or measure) the superposition, one that involves but a single step. This is to treat each data item within a given supracontext as having a single pointer to every data item (including itself) within that supracontext. Observation then is equivalent to randomly selecting one of those pointers and predicting the outcome in accord with the outcome being pointed to:

| supracontext **S** | 110 | 011 | 100 | 001 |
|---|---|---|---|---|
| data items with nonzero amplitude | 110 0<br>110 1 | 011 1 | 110 0<br>110 1 | 001 1<br>011 1 |
| pointers | 110 0 → 110 **0**<br>110 0 → 110 **1**<br>110 1 → 110 **0**<br>110 1 → 110 **1** | 011 1 → 011 **1** | 110 0 → 110 **0**<br>110 0 → 110 **1**<br>110 1 → 110 **0**<br>110 1 → 110 **1** | 001 1 → 001 **1**<br>001 1 → 011 **1**<br>011 1 → 001 **1**<br>011 1 → 011 **1** |
| outcome 0 | 2/13 | 0/13 | 2/13 | 0/13 |
| outcome 1 | 2/13 | 1/13 | 2/13 | 4/13 |

In other words, we select just one pointer anywhere in the superposition. Also note that the system of pointers is what defines the uncertainty of the system and serves as the statistical basis for determining heterogeneity.

## 2.9 *Predictions Based on Other Properties*

We observe here that we can use reversible programming along with superpositioning of supracontexts to (1) identify which supracontexts have a given property, then (2) set the amplitude of each of those supracontexts with the desired property to its frequency but allow the amplitude of all the supracontexts which fail to have that property to remain at zero, and finally (3) observe the system. As before, the qubits are restricted to zeros and ones in order to allow copying. The property of Analogical Modeling is homogeneity; that is, the homogeneous supracontexts are either nonplural in outcomes or nonplural in subcontexts (or both). In other words, the property of homogeneity is a disjunction of two simpler properties. Obviously, we could make predictions based on either of those properties alone:

THE SUPRACONTEXTS ARE NONPLURAL IN OUTCOMES

| supracontext | nonplurality of outcomes | amplitude x | amplitude y | amplitude squared x | amplitude squared y |
|---|---|---|---|---|---|
| o  m  a | yes | 0 | 0 | 0 | 0 |
| o  m  - | no |   |   |   |   |
| o  -  a | yes | 0 | 0 | 0 | 0 |
| -  m  a | yes | 1 | 0 | 1 | 0 |
| o  -  - | no |   |   |   |   |
| -  m  - | no |   |   |   |   |
| -  -  a | yes | 2 | 0 | 4 | 0 |
| -  -  - | no |   |   |   |   |
|   | *predicted chances* |   |   | 5 | 0 |

THE SUPRACONTEXTS ARE NONPLURAL IN SUBCONTEXTS

| supracontext | nonplurality of subcontexts | amplitude x | amplitude y | amplitude squared x | amplitude squared y |
|---|---|---|---|---|---|
| o  m  a | yes | 0 | 0 | 0 | 0 |
| o  m  - | yes | 1 | 1 | 2 | 2 |
| o  -  a | yes | 0 | 0 | 0 | 0 |
| -  m  a | yes | 1 | 0 | 1 | 0 |
| o  -  - | yes | 1 | 1 | 2 | 2 |
| -  m  - | no |   |   |   |   |
| -  -  a | no |   |   |   |   |
| -  -  - | no |   |   |   |   |
|   | *predicted chances* |   |   | 5 | 4 |

Immediately after assigning the data items to the supracontexts, we have the most general case. In other words, if we measured the system immediately after determining the containment vector $\mathbb{C}$, we would get the following results:

NO PROPERTY IDENTIFIED FOR THE SUPRACONTEXTS

| supracontext | the supracontext is a supracontext | amplitude x | amplitude y | amplitude squared x | amplitude squared y |
|---|---|---|---|---|---|
| o  m  a | yes | 0 | 0 | 0 | 0 |
| o  m  - | yes | 1 | 1 | 2 | 2 |
| o  -  a | yes | 0 | 0 | 0 | 0 |
| -  m  a | yes | 1 | 0 | 1 | 0 |
| o  -  - | yes | 1 | 1 | 2 | 2 |
| -  m  - | yes | 3 | 1 | 12 | 4 |
| -  -  a | yes | 2 | 0 | 4 | 0 |
| -  -  - | yes | 5 | 1 | 30 | 6 |
|   | *predicted chances* |   |   | 51 | 14 |

Of course, this is equivalent to the property that the supracontext has at last one data item.

Besides looking for the properties of nonplurality in outcomes and subcontexts, we could look for any other conceivable property and use it to predict the outcome, such as in the following two examples:

THE SUPRACONTEXT CONTAINS A SINGLE DATA ITEM

| supracontext | | | the supracontext is a singleton | amplitude | | amplitude squared | |
|---|---|---|---|---|---|---|---|
| | | | | x | y | x | y |
| o | m | a | no | | | | |
| o | m | - | no | | | | |
| o | - | a | no | | | | |
| - | m | a | yes | 1 | 0 | 1 | 0 |
| o | - | - | no | | | | |
| - | m | - | no | | | | |
| - | - | a | no | | | | |
| - | - | - | no | | | | |
| | | | *predicted chances* | | | 1 | 0 |

THE SUPRACONTEXT DIFFERS FROM THE GIVEN CONTEXT BY ONLY ONE VARIABLE

| supracontext | | | one variable away from the given context | amplitude | | amplitude squared | |
|---|---|---|---|---|---|---|---|
| | | | | x | y | x | y |
| o | m | a | no | | | | |
| o | m | - | yes | 1 | 1 | 2 | 2 |
| o | - | a | yes | 0 | 0 | 0 | 0 |
| - | m | a | yes | 1 | 0 | 1 | 0 |
| o | - | - | no | | | | |
| - | m | - | no | | | | |
| - | - | a | no | | | | |
| - | - | - | no | | | | |
| | | | *predicted chances* | | | 3 | 2 |

Note that some of these properties run the risk of having no supracontexts with the desired property. For instance, there may be no singleton supracontexts or all the supracontexts one variable away from the given context may be empty. In such cases, we would be unable to predict anything.

Whether any of these or any other properties will turn out to be valuable is an open issue. The specific claim of Analogical Modeling is that the property of homogeneity predicts language behavior and, more generally, the classification or categorization of behavior. As shown in part 1, homogeneity is equivalent to the property of never accepting a supracontext whose subcontextual analysis would increase its uncertainty (as measured by the number of disagreements in outcome between pairs of data items). In this paper I have not derived homogeneity directly in terms of pointers between data items, but in principle this can be done. The main point here is that general quantum computing algorithms can be devised, providing one makes sure that the reversible operators maintain states of only zero and one.

## 2.10 A General Summary

We may generalize the procedure used in QAM as follows. Given a set $\mathbb{S}$, we first assign instances to the members of $\mathbb{S}$ according to some qualifying characterization (in the case of QAM, it is contextual inclusion). The members of $\mathbb{S}$ form a superposition. While in superposition, we determine which members of $\mathbb{S}$ have a particular property $\mathcal{P}$. This property must be determined simultaneously but independently for each member in $\mathbb{S}$ and by using only classical reversible operators with qubits restricted to the states zero and one, thus allowing for copying of qubits. Each member of $\mathbb{S}$ with the property $\mathcal{P}$ is assigned an amplitude proportional to the frequency of the instances assigned to that member of $\mathbb{S}$, while all other members of $\mathbb{S}$ (those without the property $\mathcal{P}$) are assigned an amplitude of zero. Nonoccurring members of $\mathbb{S}$ (those for which no instances are assigned) will also have an amplitude of zero. Observation of the superposition proceeds in two stages. We first reduce the set $\mathbb{S}$ to a single member. The probability of selecting a member with the property $\mathcal{P}$ is proportional to the square of the frequency of the instances assigned to that member. Then, in order to predict an actual instance of that member, we randomly select one of the instances assigned to the member. In QAM, the set $\mathbb{S}$ is a power set and is in superposition; the members of the set are the supracontexts. The property $\mathcal{P}$ is homogeneity. For each member in $\mathbb{S}$, the property of homogeneity for each member in $\mathbb{S}$ can be determined in linear time (and with linear resources) – that is, without the exponential processing inherent in solving this problem using classical programming.

## 2.11 The General Algorithm

In the following, I provide the the general algorithm for QAM. Generally, $m$ represents the number of data items in the data set, $n$ the number of feature variables, and $\omega$ the number of outcome variables  In this, I do not show the derivation of the constant qubit vectors: the differences **D**, the feature variables **F**, and the outcome variables **Ω**; I will assume that they have already been provided for. (See section 2.4 for **D** and section 2.7 for **F** and **Ω**.) The two derived operators ONES and INCLUSION are defined at the end of the summary.

THE QUBITS AND THEIR INITIAL SETTINGS

    *controlling supracontextual superposition*

        $\mathbb{S}: \{0,1\}^n$

    *initial states for the substantive qubit vectors, with the number of qubits*

| | | |
|---|---|---|
| $\mathbb{C} = \mathbb{0}$ | $m$ | containment |
| $\mathbb{P} = \mathbb{1}$ | $n$ | nonplurality of feature variables |
| $\mathbb{Q} = \mathbb{1}$ | $\omega$ | nonplurality of outcome variables |
| $\mathbb{M} = \mathbb{1}$ | 1 | plurality of subcontexts |
| $\mathbb{N} = \mathbb{1}$ | 1 | plurality of outcomes |
| $\mathbb{H} = \mathbb{1}$ | 1 | homogeneity |
| $\mathbb{A} = \mathbb{0}$ | $m$ | amplitude |

*initial states for the auxiliary qubit vectors, with the number of qubits*

$\mathbb{U} = 0$     $n$
$\mathbb{V} = 0$     $n$
$\mathbb{W} = 1$     $n$
$\mathbb{Z} = 0$     $n + 1$
$\mathbb{X} = 1$     $m$
$\mathbb{Y} = 0$     $m + 1$
$\Phi = 0$     $\omega$
$\Psi = 0$     $\omega$
$\Upsilon = 0$     $\omega + 1$

THE EVOLUTION OF THE SYSTEM

*determining which data items are contained in* $\mathbb{S}$                [section 2.6]

     from $j = 1$ to $m$ do

         INCLUSION($\mathbb{S}$, $\mathbf{D}^{[j]}$, $\mathbb{W}$, $\mathbb{Z}$)
         CNOT($\mathbb{Z}_n$, $\mathbb{C}$)
         INCLUSION$^{-1}$($\mathbb{S}$, $\mathbf{D}^{[j]}$, $\mathbb{W}$, $\mathbb{Z}$)

*determining the plurality of the feature variables in* $\mathbb{S}$            [section 2.7.1]

             from $i = 1$ to $n$ do

| | |
|---|---|
| *testing for zeros* | CCNOT($\mathbb{C}$, $\mathbf{F}^{[i]}$, $\mathbb{X}$) |
| | ONES($\mathbb{X}$, $\mathbb{Y}$) |
| | CNOT($\mathbb{Y}_m$, $\mathbb{U}_i$) |
| | ONES$^{-1}$($\mathbb{X}$, $\mathbb{Y}$) |
| | CCNOT($\mathbb{C}$, $\mathbf{F}^{[i]}$, $\mathbb{X}$) |
| | |
| *testing for ones* | NOT($\mathbf{F}^{[i]}$) |
| | CCNOT($\mathbb{C}$, $\mathbf{F}^{[i]}$, $\mathbb{X}$) |
| | ONES($\mathbb{X}$, $\mathbb{Y}$) |
| | CNOT($\mathbb{Y}_m$, $\mathbb{V}_i$) |
| | ONES$^{-1}$($\mathbb{X}$, $\mathbb{Y}$) |
| | CCNOT($\mathbb{C}$, $\mathbf{F}^{[i]}$, $\mathbb{X}$) |
| | NOT($\mathbf{F}^{[i]}$) |
| | |
| *all zeros or all ones?* | NOT($\mathbb{U}_i$) |
| | NOT($\mathbb{V}_i$) |
| | CCNOT($\mathbb{U}_i$, $\mathbb{V}_i$, $\mathbb{P}_i$) |

*determining the plurality of the outcome variables in* $\mathbb{S}$ [section 2.7.2]

    from $\iota = 1$ to $\omega$ do

*testing for zeros*    CCNOT($\mathbb{C}, \mathbf{\Omega}^{[\iota]}, \mathbb{X}$)
ONES($\mathbb{X}, \mathbb{Y}$)
CNOT($\mathbb{Y}_m, \mathbb{\Phi}_\iota$)
ONES$^{-1}$($\mathbb{X}, \mathbb{Y}$)
CCNOT($\mathbb{C}, \mathbf{\Omega}^{[\iota]}, \mathbb{X}$)

*testing for ones*    NOT($\mathbf{\Omega}^{[\iota]}$)
CCNOT($\mathbb{C}, \mathbf{\Omega}^{[\iota]}, \mathbb{X}$)
ONES($\mathbb{X}, \mathbb{Y}$)
CNOT($\mathbb{Y}_m, \mathbb{\Psi}_\iota$)
ONES$^{-1}$($\mathbb{X}, \mathbb{Y}$)
CCNOT($\mathbb{C}, \mathbf{\Omega}^{[\iota]}, \mathbb{X}$)
NOT($\mathbf{\Omega}^{[\iota]}$)

*all zeros or all ones?*    NOT($\mathbb{\Phi}_\iota$)
NOT($\mathbb{\Psi}_\iota$)
CCNOT($\mathbb{\Phi}_\iota, \mathbb{\Psi}_\iota, \mathbb{Q}_\iota$)

*determining the plurality of the subcontexts in* $\mathbb{S}$ [section 2.7.1]

    ONES($\mathbb{P}, \mathbb{Z}$)
    CNOT($\mathbb{Z}_n, \mathbb{M}$)
    ONES$^{-1}$($\mathbb{P}, \mathbb{Z}$)

*determining the plurality of the outcomes in* $\mathbb{S}$ [section 2.7.2]

    ONES($\mathbb{Q}, \mathbb{\Upsilon}$)
    CNOT($\mathbb{\Upsilon}_\omega, \mathbb{N}$)
    ONES$^{-1}$($\mathbb{Q}, \mathbb{\Upsilon}$)

*determining the homogeneity in* $\mathbb{S}$ [section 2.7.3]

    CCNOT($\mathbb{M}, \mathbb{N}, \mathbb{H}$)

*determining the amplitude in* $\mathbb{S}$ [section 2.8]

    from $j = 1$ to $m$ do
        CCNOT($\mathbb{C}_j, \mathbb{H}, \mathbb{A}_j$)

OBSERVATION (using pointers)

    Randomly select a pointer to any data item in the (superpositioned) amplitude $\mathbb{A}$;
    the predicted behavior is the outcome associated with that data item.

    ONES($\mathbb{X}$, $\mathbb{Y} = \mathbb{0}$), with $\mathbb{X}$ of length *L* and $\mathbb{Y}$ of length *L* + 1         [section 2.3.2]

        NOT($\mathbb{Y}_0$)
        from *k* = 1 to *L* do
            CCNOT($\mathbb{X}_k$, $\mathbb{Y}_{k-1}$, $\mathbb{Y}_k$)

    INCLUSION($\mathbb{S}$, **D**, $\mathbb{W} = \mathbb{1}$, $\mathbb{Z} = \mathbb{0}$)         [section 2.6]

        CCNOT($\mathbb{S}$, **D**, $\mathbb{W} = \mathbb{1}$)
        ONES($\mathbb{W}$, $\mathbb{Z} = \mathbb{0}$)

*Acknowledgments*

I wish to thank my colleagues in the Analogical Modeling Research Group for providing useful input in preparing this paper. I especially wish to thank Don Chapman and Deryle Lonsdale for their careful reading of the text of this paper. Theron Stanford has provided helpful suggestions on the representations.